\documentclass[%
reprint,
amsmath,amssymb,
aps,twocolumn
]{revtex4-2}
\usepackage{graphicx}

\usepackage{subcaption}
\usepackage{caption}
\usepackage{epsfig}
\usepackage{float}
\usepackage{graphicx}
\usepackage{amsmath}
\usepackage{dcolumn}
\usepackage{bm}
\usepackage[pagebackref=false,colorlinks=true,linkcolor=blue,citecolor=red,urlcolor=blue]{hyperref}
\usepackage[english]{babel}
\usepackage{xcolor}
\makeatletter
\renewcommand*\l@section{\@dottedtocline{0}{1em}{2.3em}}
\renewcommand*\l@subsection{\@dottedtocline{0}{1em}{2.3em}}
\makeatother
\begin{document}
	
	\preprint{APS/123-QED}
	
	\title{Accelerating FJNW Metric}
	
	\author{Homayon Anjomshoa$^{1}$}
	\email{h.anjomshoa@ph.iut.ac.ir}
	\author{Behrouz Mirza$^{1}$   }
	\email{b.mirza@iut.ac.ir}
	\author{Alireza
		Azizallahi$^{1}$}
	\email{azizallahialireza@gmail.com}
	
	\affiliation{$^{1}$ Department of Physics, Isfahan University of Technology, Isfahan 84156-83111, Iran}

	
	\date{\today}

	\begin{abstract}
		We derive exact form of accelerating Fisher-Janis-Newman-Winicour (FJNW) metric by a simple  perturbative method.  We also argue that by using Buchdahl
		transformations one  can  obtain the same accelerating FJNW metric. We investigate 
		singularities of the accelerating FJNW metric and study their effects on global and local structures of this spacetime. We also study geodesics and stability of circular orbits.
		\\
		
		\begin{description}
			\item[PACS numbers]C-metric, Accelerating FJNW, Stability of circular orbits.
			
			\tableofcontents
			
		\end{description}
	\end{abstract}
	
	\pacs{Valid PACS appear here}
	\maketitle

	
	\section{\label{sec:level1}Introduction}
	
	The first model for a physically possible but highly idealized system that undergoes gravitational collapse was introduced by Oppenheimer and Snyder in 1939 \cite{oppenhimer.1939}. This model was generally overlooked by physics community at the time of its publication. This disregard was mainly due to the prediction of singularity in this model, a region of spacetime that physical observables like density and curvature grow unbounded. At that time the scientific zeitgeist considered these singularities coincidences stumbled upon the theory of general relativity (GR) due to the non-realistic assumptions, such as spherical symmetry which is not perfectly attainable in real world and absence of pressure in the model. But there was a paradigm shift during 1960s. Roger Penrose has argued that singularities in GR are far more common and fundamental in GR and contrary to what previously thought they cannot be simply attributed to idealization and approximation in our models. In this theorem Penrose goes roughly the following line of arguments. He takes the spacetime manifold to be Lorentzian and time-orientable one which has a Cauchy surface and a trapped surface of co-dimension 2. He then proved, that these assumptions leads to existence of singularities \cite{penrose.1965}. It is worth mentioning that  what is actually proved is geodesic incompleteness of the spacetime and this is generally different from physical parameters such as density or pressure divergency. Due to this fact it is generally accepted that geodesic incompleteness  deserved to be considered singular feature of spacetime. However, there is no consensus over definition of singularity in GR. This result could be used to show that a gravitational collapse in GR quite generally results in singular spacetimes. After this breakthrough  several people; such as Hawking, Geroch and Penrose himself, generalized these results to other cases  that could be used for different scenarios such as cosmological one  \cite{penrose.1969,geroch.1966,geroch.1967,geroch.1968,hawking.1965,hawking.1966,hawking.1966b,hawking.1966c,hawking.2014}                These theorems demonstrated that singularities in GR are common,  especially in specific physically interesting scenarios such as gravitational collapse models. Geodesic incompleteness was sufficiently  problematic that motivated Penrose to propose a conjecture that these singular solutions of GR are avoided by nature in generic cases \cite{penrose.1999,penrose.1969}. There are two versions of this conjecture, weak cosmic censorship conjecture (CCC) and strong CCC. Weak version claims in generic solutions of GR singularities are always covered by an event horizon. On the other hand, strong version intends a more ambitious statement by claiming impossibility of generic extensions of globally hyperbolic solutions of GR passing Cauchy horizon. Therefore in essence weak cosmic censorship conjecture (WCCC) is about observability of singularities; however, strong cosmic censorship conjecture (SCCC) deals with predictability of generic solutions of GR. Whether or not these conjectures are correct is controversial, and there is no formal proof or disproof of them in general cases. Some research suggest that the ultimate fate of systems of non-exotic matters and fields with reasonable initial conditions could lead to the formation of naked singularities as well as black-holes \cite{singolar.1,singolar.2,singolar.3,singolar.4,singolar.5,singolar.6,singolar.7,singolar.8,singolar.9,singolar.10,singolar.11,singolar.12,singolar.13,singolar.14,singolar.15,singolar.16}. However, these results challenge a rigorous formulation of CCC, we should be cautious not to interpret them as a definite violation of CCC. \\
	In this paper we introduce a class of exact solutions of Einstein equations in the presence of a scalar field that belongs to the Weyl class of solutions with axial symmetry. The class of solutions have three free parameters; one of them is the mass of the source, the second one is due to the presence of the scalar field and the third one can be interpreted as acceleration. The class of metrics represent accelerating form of FJNW metric and reduces to FJNW  and C-metric at certain values of the parameters.  \\
	C-metric is a widely known solution to Einstein equation that belongs to the weyl class that appeared in the work of Weyl for the first time \cite{weylcmetric.1917}. This metric was rediscovered by Newman Tamburino in 1960s \cite{newmancmetric.1961}. The name of this metric come from work of Ehlers and Kundt \cite{Lwitten} and the solution also appeared in the work of Kimmersley as one of the cases of all Petrov type D solutions \cite{dtipcmetric}. Explanation about casual structure of this metric can be found in \cite{causalcmetric1,causalcmetric2,causalcmetric3}. The shadow of C-metric was explored in \cite{shadowcmetric1,shadowcmetric2,shadowcmetric3,shadowcmetric4,shadowcmetric5,shadowcmetric6}. Radiative properties of this space-time are studied in \cite{reffff1}.  C-metric in asymptotically ADS spacetime can be found in \cite{adscmetric}. 
	In this paper we study a class of C-metrics in the presence of a scalar field.\\
	 This paper is organized as follows.
	In Section II we introduce a class of accelerating metrics derived through a simple perturbative approach and represent it in different coordinate systems. In Section III, we analyze the admissible coordinate ranges and identify the singularities associated with the geometry. Section IV is devoted to computing the effective potential, exploring the structure of stable circular orbits, and examining the motion of test particles in this spacetime.

	\section{\label{II}Accelerating FJNW Metric}
	FJNW metric, that introduced by Fisher \cite{FJNW} and Janis, Newman and Winicour in \cite{JNWorg}, is a class of exact solutions of the Einstein's equations in the presence of a scalar field. Recently, a new class of solutions in the presence of a scalar field has been derived \cite{ostad1,ostad2,ostad3,ostad4,ostad6,ostad5,ostad7}.\\
	In this Section we derive the accelerating form of  FJNW metric that in certain limits reduces to the FJNW and C-metrics. \\
	The signature of the metric is assumed to be $\left(-,+,+,+\right)$ and $c=G=1$. We consider the following action,
	\begin{equation}
		\label{actionnew}
		\begin{split}
			&\mathcal{S}=\int dx^{4}  \sqrt{g}  \left(R-\partial_{\mu}\phi\left(r,\theta\right)\partial^{\mu}\phi\left(r,\theta\right)\right)
		\end{split},
	\end{equation} 
	where, $R$ and $\phi\left(r,\theta\right)$ are the Ricci scalar and the scalar field respectively and $g$ is the determinant of the metric. Equations of motion are obtained as follows:

	\begin{equation}
		\label{R-Phi}
		\begin{split}
			&R_{\mu\nu}=\partial_{\mu}\phi\left(r,\theta\right)\partial_{\nu}\phi\left(r,\theta\right),
		\end{split}
	\end{equation}
and
	\begin{equation}
		\label{equation-scaler}
		\begin{split}
			&\square \phi(r,\theta)  =0.
		\end{split}
	\end{equation}
The FJNW metric can be written as follows \cite{JNWgood}:
	\begin{equation}
		\label{metricjnw}
		\begin{split}
			ds^2=&-f(r)^{\lambda}dt^2+f(r)^{-\lambda}dr^2\\
			&+f(r)^{1-\lambda}r^2\left(d\theta^2+\sin\left(\theta\right)^2d\varphi^2\right),\\
			\phi\left(r\right)=&\sqrt{\dfrac{1-\lambda^2}{2}}\ln\left(f(r)\right),
		\end{split}
	\end{equation}
where,
	\begin{equation}
		\label{fjnw}
		\begin{split}
			f(r)=1-\dfrac{2m}{r}.
		\end{split}
	\end{equation}
	Now we want to find the accelerating form of FJNW metric. In order to obtain a new solution with the desired properties, a metric in the following general form, is assumed,
	\begin{equation}
		\label{metric1}
		\begin{aligned}
			&ds^{2}=\dfrac{f\left(r\right)^{\lambda}h\left(r\right)^{a_{1}}}{\Omega\left(r,\theta\right)^{2a_{2}}}dt^{2}+\dfrac{1}{\Omega\left(r,\theta\right)^{a_{3}}}\left[\ \dfrac{1}{f\left(r\right)^{\lambda}h\left(r\right)^{a_{1}}}dr^{2}
			\right.\\
			&\left. f\left(r\right)^{1-\lambda}h\left(r\right)^{a_{4}}\times r^{2}\left(\dfrac{d\theta^{2}}{\mathcal{G}\left(\theta\right)^{a_{5}}}+\mathcal{G}\left(\theta\right)^{a_{5}}\sin\left(\theta\right)^{2}d\varphi^{2}\right)\right],\\ 
		\end{aligned}
	\end{equation}
	where,
	\begin{equation}
		\label{laps1}
		\begin{split}
			&f\left(r\right)=1-\dfrac{2m}{r}, \\
			&h\left(r\right)=1-a^{2}r^{2},\\
			&\Omega\left(r,\theta\right)=1+ar\cos\left(\theta\right),\\
			&\mathcal{G}\left(\theta\right)=1+2ma\cos\left(\theta\right),		
		\end{split}
	\end{equation}
	here $m$ is proportional to mass and $a$ represents the acceleration. It turns out that the solution to Eq.~\eqref{equation-scaler}, similar to the case of FJNW, is proportional to the logarithm of $g_{tt}$ in the following form, 
	
	\begin{equation}
		\label{scaler-field1}
		\begin{split}
			&\phi\left(r,\theta\right)=C_{1} \times \ln\left(\dfrac{f\left(r\right)h\left(r\right)}{\Omega\left(r,\theta\right)^{2}}\right).
		\end{split}
	\end{equation}

	By assuming that the scalar field $\varphi\left(r,\theta\right)$ is independent of $t$ and $\phi$, Eq.~\eqref{R-Phi} for $R_{tt}$ and $R_{\varphi \varphi}$ reduces to $R_{tt}=0$ and $R_{\varphi \varphi}=0$. Now by using metric in Eq.~\eqref{metric1} and expansion of  $R_{tt}$ and $R_{\varphi \varphi}$ up to second orders of $a$ and $m$ we determine the free parameters ($a_{1},a_{2},a_{3},a_{4},a_{5}$). The expansion for $R_{tt}$ to the first order of $a$ is
	
	\begin{equation}
		\label{rtt1}
		\begin{split}
			R_{tt}  \propto &-\dfrac{\cos\left(\theta\right)\left(\left(a_{2}+\dfrac{a_{3}}{2}\right)\lambda-2a_{2}\right)m a}{r^{2}}
			+\mathcal{O}\left(a^{2}\right),
		\end{split}
	\end{equation}
	Therefore $R_{tt}=0$ yields,
	\begin{equation}
		\label{a3}
		\begin{split}
			a_{3}=\dfrac{2 a_{2}\left(2-\lambda\right)}{\lambda}.
		\end{split}
	\end{equation}
	Now  by substitution of $a_{3}$ in metric in Eq.~\eqref{metric1} and expanding $R_{tt}$ to the second order of $a$ we have
	\begin{equation}
		\label{rtt1}
		\begin{split}
			R_{tt}&\propto-\dfrac{-3 a^{2}\left(\left(a_{1}-\dfrac{a_{2}}{3}\right)\lambda-\dfrac{a_{2}}{2}\right)}{\lambda}
			+\mathcal{O}\left(a^{3}\right),
		\end{split}
	\end{equation}
	Using $R_{tt}=0$, one has,
	\begin{equation}
		\label{a1}
		\begin{split}
			&a_{1}=\dfrac{a_{2}\left(\lambda+2a_{3}\right)}{3\lambda}.
		\end{split}
	\end{equation}
	Now again by the substitution of $a_{1}$ in metric in Eq.\eqref{metric1} and expansion of $R_{\varphi\varphi}$ to the lowest order of $m$ we obtain
	\begin{equation}
		\label{rphiphi1}
		\begin{split}
			&R_{\varphi\varphi}\propto\dfrac{6 m a \sin\left(\theta\right)^2\cos\left(\theta\right)\left(a_{5}\lambda-a_{2}\right)}{\lambda}+\mathcal{O}\left(a^{2}\right),
		\end{split}
	\end{equation}
	Using $R_{\varphi\varphi}=0$, one has,
	\begin{equation}
		\label{a5}
		\begin{split}
			&a_{5}=\dfrac{a_{2}}{\lambda}.
		\end{split}
	\end{equation}
	Substituting $a_{5}$ in Eq.\eqref{metric1} and expanding $R_{\varphi\varphi}$ to the second order of $a$ we obtain
	\begin{equation}
		\label{Rphiphi}
		\begin{split}
			R_{\varphi\varphi}\propto&\dfrac{a^{2}\sin\left(\theta\right)^{2}\left(2a_{2}^{2}\lambda+a_{2}\lambda^{2}+3a_{4}\lambda^{2}-2a_{2}^{2}-a_{2}\lambda\right)}{\lambda^{2}}\\
			&+\mathcal{O}\left(a^{3}\right),
		\end{split}
	\end{equation}
	Using $R_{\varphi\varphi}=0$ one has,
	\begin{equation}
		\label{a4}
		\begin{split}
			&a_{4}=\dfrac{a_{2}\left(2a_{2}+\lambda\right)\left(\lambda-1\right)}{3\lambda^{2}}.
		\end{split}
	\end{equation}
	Now we have $a_{1}$, $a_{3}$, $a_{4}$ and $a_{5}$ as a function of $a_{2}$. In order to find $a_{2}$, we expand $R_{\varphi\varphi}$ to the third order of $a$ and obtain
	\begin{equation}
		\label{Rphiphi3}
		\begin{split}
			R_{\varphi\varphi}\propto&\dfrac{2a^{3}\sin\left(\theta\right)^{2}\cos\left(\theta\right)\left(a_{2}-\lambda\right)}{\lambda^{3}}+\mathcal{O}\left(a^{4}\right),
		\end{split}
	\end{equation}
	Therefore we derive all parameters as follows,
	
	\begin{equation}
		\label{a}
		\begin{split}
			&a_{1}=\lambda,\\
			&a_{2}=\lambda,\\
			&a_{3}=2\left(2-\lambda\right),\\
			&a_{4}=1-\lambda,\\
			&a_{5}=1.
		\end{split}
	\end{equation}
	
	Now by putting determined parameters in Eq.\eqref{a} into Eq.\eqref{metric1}, the final form of the metric is written as follows:
	\begin{equation}
		\label{metric2}
		\begin{aligned}
			&ds^{2}=-\dfrac{f\left(r\right)^{\lambda}h\left(r\right)^{\lambda}}{\Omega\left(r,\theta\right)^{2\lambda}}dt^{2}+\dfrac{1}{\Omega\left(r,\theta\right)^{2\left(2-\lambda\right)}}\left[\ \dfrac{dr^{2}}{f\left(r\right)^{\lambda}h\left(r\right)^{\lambda}}
			\right.\\
			&\left. f\left(r\right)^{1-\lambda}h\left(r\right)^{1-\lambda}\times r^{2}\left(\dfrac{d\theta^{2}}{\mathcal{G}\left(\theta\right)}+\mathcal{G}\left(\theta\right)\sin\left(\theta\right)^{2}d\varphi^{2}\right)\right].\\ 
		\end{aligned}
	\end{equation}
	It should be noted that the metric in Eq.\eqref{metric2} is an exact solution of the Einstein-scalar theory. It is interesting that a simple perturbative method leads to an exact solution of the Einstein field equations. The scalar field in Eq.\eqref{scaler-field1} has an undetermined constant. We find this constant by solving the following set of equations,
	\begin{equation}
		\label{R-matrix}
		\begin{split}
			R_{\mu \nu}=
			\begin{pmatrix}
				0&0&0&0\\0&\left(\partial_{_{r}}\phi\right)^{2}&\left(\partial_{_{r}}\phi\right)\left(\partial_{_{\theta}}\phi\right)&0\\0&\left(\partial_{_{\theta}}\phi\right)\left(\partial_{_{r}}\phi\right)&\left(\partial_{_{\theta}}\phi\right)^{2}&0\\0&0&0&0
			\end{pmatrix}.
		\end{split}
	\end{equation}
Now, by considering Eq.~\eqref{scaler-field1} and Eq.~\eqref{metric2} and simplifying $R_{rr}=\left(\partial_{_{r}}\phi\right)^{2}$, the constant coefficient of the scalar field can be derived as follows:
	\begin{equation}
		\label{21}
		\begin{split}
			&C_{1}=\sqrt{\dfrac{1-\lambda^2}{2}}.
		\end{split}
	\end{equation}

\begin{figure}[H]
		\centering
		\captionsetup{justification=justified,singlelinecheck=false}
		\begin{subfigure}{0.65\linewidth}
			\includegraphics[width=\linewidth]{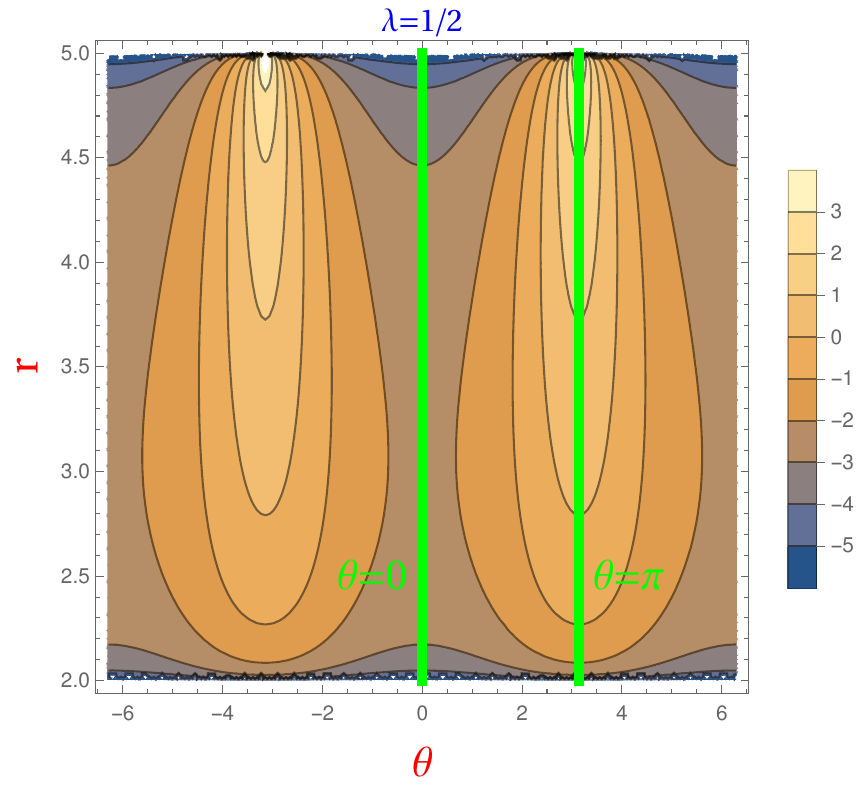}
			\caption{\it $\lambda$ is constant.}
			\label{fig:subfig1}
		\end{subfigure}
		\begin{subfigure}{0.65\linewidth}
			\includegraphics[width=\linewidth]{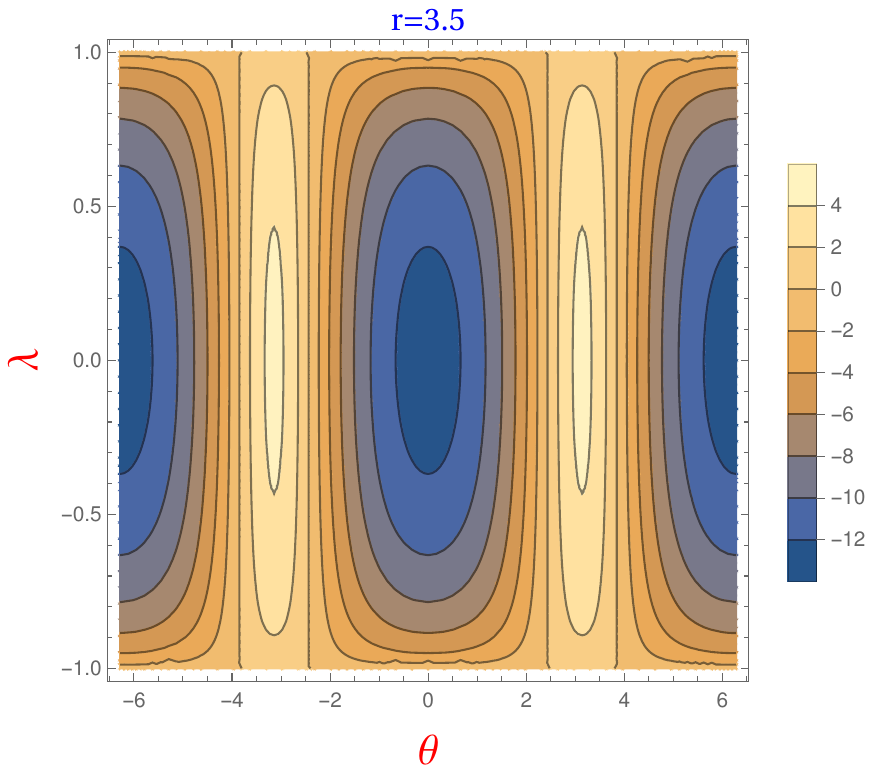}
			\caption{\it $r$ is constant.}
			\label{fig:subfig2}
		\end{subfigure}
		
		\captionsetup{justification=justified}
		\caption[]{\itshape Variations of the scalar field Eq.\eqref{q} based on the parameters $\lambda$, $r$ and $\theta$, where $m=1$ and $a=\dfrac{1}{5}$. 
			In diagram (a), the value of $\lambda$ is held constant, and the variations of the scalar field are plotted as a function of $r$ and $\theta$. In diagram (b), r is held constant while examining the behavior of the scalar field. Note that, based on the chosen coordinate system, only the region between the two green lines is physically meaningful.}
		\label{fig:subfigures1}
	\end{figure}

It should be noted that one can also use $R_{r\theta}=\left(\partial_{_{r}}\phi\right)\left(\partial_{_{\theta}}\phi\right)$ or $R_{\theta\theta}=\left(\partial_{_{\theta}}\phi\right)^{2}$ to obtain $C_{1}$. Now, by substituting Eq.~\eqref{21} into Eq.~\eqref{scaler-field1}, the general form of the scalar field is written as follows:
	\begin{equation}
		\label{q}
		\phi(r,\theta)=\sqrt{\dfrac{1-\lambda^2}{2}}\ln\left(\dfrac{\left(1-\dfrac{2m}{r}\right)\left(1-a^{2}r^{2}\right)}{\left(1+ar\cos\left(\theta\right)\right)^{2}}\right).
	\end{equation}
	We assume that the allowed range of $\lambda$ is $-1<\lambda<1$, to have a real scalar field.\\
	It is interesting that  divergences of the scalar field corresponds to that of the accelerating metric. The  Ricci  and Kretschmann scalars are derived in the following Sections. The scalar field in Eq.~\eqref{q} is depicted in FIG.~(\ref{fig:subfigures1}). 

	\subsection{Buchdahl transformations}
	Buchdahl theorem makes one able to derive new solutions of Einstein equations from already existing vacuum solutions of these equations \cite{BB1,BB2}. Second form of the theorem says if the following d-dimensional static metric,
	\begin{equation}
		\label{Buchdahlmetric}
		\begin{split}
			ds^{2}&=g_{\alpha\beta}dx^{\alpha}dx^{\beta}=g_{tt}\left(dx^{t}\right)^{2}+g_{ij}dx^{i}dx^{j}
		\end{split},
	\end{equation}
	be a vacuum solution of Einstein equations, the configuration 
	\begin{equation}
		\label{Buchdahlmetric1}
		\begin{split}
			ds_{B}^{2}&=\left(g_{tt}\right)^{\lambda}\left(dx^{t}\right)^{2}+\left(g_{tt}\right)^{1-\lambda}g_{ij}dx^{i}dx^{j},\\
			\phi_{B}&=\sqrt{\dfrac{1-\lambda^{2}}{2}}\ln\left(g_{tt}\right),
		\end{split}
	\end{equation}
	is also a solution of
	\begin{equation}
		\label{Buchdahlmetric2}
		\begin{split}
			R_{\mu\nu}&=\partial_{\mu}\phi_{B}\partial_{\nu}\phi_{B},\\
			\square \phi_{B} &=0.
		\end{split}
	\end{equation}
	If we apply this theorem to the following C-metric:
	\begin{equation}
		\label{Buchdahlmetric3}
		\begin{split}
			ds^{2}=&\dfrac{f\left(r\right)h\left(r\right)}{\Omega\left(r,\theta\right)^{2}}dt^{2}+\dfrac{1}{\Omega\left(r,\theta\right)^{2}}\left[\ \dfrac{1}{f\left(r\right)h\left(r\right)}dr^{2}
			\right.\\
			&\left.  r^{2}\left(\dfrac{d\theta^{2}}{\mathcal{G}\left(\theta\right)}+\mathcal{G}\left(\theta\right)\sin\left(\theta\right)^{2}d\varphi^{2}\right)\right],
		\end{split}
	\end{equation}
	where
	\begin{equation}
		\label{Buchdahlmetric4}
		\begin{split}
			f\left(r\right)&=1-\dfrac{2m}{r}, \\
			h\left(r\right)&=1-a^{2}r^{2},\\
			\Omega\left(r,\theta\right)&=1+ar\cos\left(\theta\right),\\
			\mathcal{G}\left(\theta\right)&=1+2ma\cos\left(\theta\right).		
		\end{split}
	\end{equation}
	We reach to the metric
	\begin{equation}
		\label{metricB2}
		\begin{aligned}
		&ds^{2}=\dfrac{-f\left(r\right)^{\lambda}h\left(r\right)^{\lambda}}{\Omega\left(r,\theta\right)^{2\lambda}}dt^{2}+\dfrac{1}{\Omega\left(r,\theta\right)^{2\left(2-\lambda\right)}}\left[\ \dfrac{dr^{2}}{f\left(r\right)^{\lambda}h\left(r\right)^{\lambda}}
			\right.\\		
		&\left. +f\left(r\right)^{1-\lambda}h\left(r\right)^{1-\lambda}\times r^{2}\left(\dfrac{d\theta^{2}}{\mathcal{G}\left(\theta\right)}+\mathcal{G}\left(\theta\right)\sin\left(\theta\right)^{2}d\varphi^{2}\right)\right],		
				\end{aligned}
	\end{equation}
	
	This class of metrics are exact solutions of Einstein-scalar field equations with
	\begin{equation}
		\label{metricB2}
		\begin{split}
			&\phi=\sqrt{\dfrac{1-\lambda^2}{2}}\ln\left(\dfrac{\left(1-\dfrac{2m}{r}\right)\left(1-a^{2}r^{2}\right)}{\left(1+ar\cos\left(\theta\right)\right)^{2}}\right). 
		\end{split}
	\end{equation}
	This metric is exactly equivalent to the metric in Eq.~\eqref{metric2} obtained in section \ref{II}.
	
	\subsection{Weyl coordinates}
	
	In this Section, we briefly show that the metric obtained in Section \ref{II} can also be written in the following form, using a combination of hyperbolic transformations and Weyl coordinates. The general expression of the Weyl metric is given below:
	
	\begin{equation}
		\label{WeylU0}
		\begin{split}
			&ds^{2}=-e^{2U}dt^{2}+e^{-2U}\left(e^{2\gamma}\left(d\rho^{2}+dz^{2}\right)+\rho^{2}d\Phi^{2}\right).
		\end{split}
	\end{equation}
	We consider the following coordinate transformations
	\begin{equation}
		\label{WeylU2}
		\begin{aligned}
			&\rho=r\sin\left(\theta\right)\dfrac{\sqrt{f\left(r\right)h\left(r\right)\mathcal{G}\left(\theta\right)}}{\Omega\left(r,\theta\right)^{2}},\\
			&z=\frac{\left(r-m\right)\cos\left(\theta\right)+ar\left(r-m\left(\sin\left(\theta\right)^{2}-ar\cos\left(\theta\right)\right)\right)}{\Omega\left(r,\theta\right)^2}.
			 \\
		\end{aligned}
	\end{equation}\\
	If we define unknown functions in the general form of the  Weyl metric based on coordinates $r$ and $\theta$ in the following form:
	\begin{equation}
		\label{WeylU1}
		\begin{split}
			e^{2U}=&\left(\dfrac{R_{1}+R_{2}-2m}{R_{1}+R_{2}+2m}\right)^{\lambda}\left(\dfrac{h\left(r\right)	\mathcal{G}\left(\theta\right)}{\Omega\left(r,\theta\right)^{2}}\right)^{\lambda},\\
			e^{2\gamma}=&\left(\dfrac{\left(R_{1}+R_{2}\right)^{2}-4m}{4R_{1}R_{2}}\right)\times\\
			&\left(\dfrac{h(r)}{\Omega\left(r,\theta\right)\mathcal{G}\left(\theta\right)\left(1-2a^{2}mr+a\left(2m-r\right)\cos\left(\theta\right)\right)}\right),
		\end{split}
	\end{equation}
	where
	\begin{equation}
		\label{WeylU2}
		\begin{split}
			R_{1}=&\dfrac{r-m\left(1+\cos\left(\theta\right)\right)-mar\left(1-\cos\left(\theta\right)\right)}{\Omega\left(r,\theta\right)},\\
			R_{2}=&\dfrac{r-m\left(1-\cos\left(\theta\right)\right)+mar\left(1+\cos\left(\theta\right)\right)}{\Omega\left(r,\theta\right)},
		\end{split}
	\end{equation}
	and perform the coordinate transformation on the Weyl metric with the functions $e^{2U}$ and $e^{2\gamma}$ as defined in Eq.~\eqref{WeylU1}, we arrive at the same expression in Eq.~\eqref{metric2} of section \ref{II}.
	
	\subsection{\label{sec:level2}Accelerating FJNW metric in $\left(t,X,Y,\varphi\right)$ coordinates}
	In 1917 the C-metric was derived by Luigi Levi-Civita. \cite{causalcmetric1}. It was mainly disregarded for an extended period, however, it attracted attention of physicists in the next decades and major development appeared \cite{cmetricatentio1,cmetricatentio2,cmetricatentio3}.\\
	The general form of C-metric can be written as follows \cite{C-metric}.
	\begin{equation}
		\label{C-metric1}
		\begin{split}
			&ds^2=\dfrac{1}{a^{2}\left(X+Y\right)^{2}}\times \\ &\left(-F(Y)dt^{2}+\frac{dY^{2}}{F(Y)}+\frac{dX^{2}}{G(X)}+G(X)d\varphi^{2}\right).
		\end{split}
	\end{equation}
	In Eq. \eqref{C-metric1}, $F(Y)$ and $G(X)$ are functions of the third order of $X$ and $Y$ and defined as follows:
	
	\begin{equation}
		\label{C-metricfanction}
		\begin{split}
			&F(Y)=-1+Y^{2}-2maY^{3},\\
			&G(X)=1-X^{2}-2maX^{3},
		\end{split}
	\end{equation}
where the parameter $m$ denotes mass. If we set $m=0$ and apply appropriate coordinate transformations, the metric turns into the Rindler metric, implying that the parameter $a$ can be understood as acceleration. 
	Comparing the two relations, in Eq. \eqref{C-metricfanction}, we have  $F(\omega)=-G(-\omega)$. If condition $27m^{2}a^{2}<1$ is satisfied, these two functions have three different roots. If this condition is established, this metric will yield interesting results \cite{C-metric}.
Now, by using the following coordinate transformation $Y=-1$, $X=cos\left(\theta\right)$ and $\tau=at$	in Eq.~\eqref{metric2}, one can derive the accelerating FJNW metric as below\\
	\begin{equation}
		\label{XYmetric}
		\begin{split}
			ds^{2}&=-\dfrac{\mathcal{F}\left(Y\right)^{\lambda}}{a^{2}\left(X+Y\right)^{2\lambda}}d\tau^{2}+ \\ &\dfrac{1}{a^{2}\left(X+Y\right)^{2\left(2-\lambda\right)}}\left[\dfrac{dY^{2}}{\mathcal{F}\left(Y\right)^{\lambda}}+ \right. \\ &\left. \mathcal{F}\left(Y\right)^{1-\lambda}\left(\dfrac{dX^{2}}{\mathcal{G}\left(X\right)}+\dfrac{d\phi^{2}}{\mathcal{G}\left(X\right)^{-1}}\right)\right],
		\end{split}
	\end{equation}
	where $\mathcal{F}(Y)$ and $\mathcal{G}(X)$ are defined in Eq.~\eqref{C-metricfanction}. \\
	The scalar field $\phi\left(X,Y\right)$, can be derived by solving Eq.~\eqref{equation-scaler} as follows:
	\begin{equation}
		\label{XYriich}
		\begin{split}
			&\phi\left(X,Y\right)=\sqrt{\dfrac{1-\lambda^{2}}{2}} \ln\left(\dfrac{\mathcal{F}\left(Y\right)}{a^{2}\left(X+Y\right)^{2}}\right).
		\end{split}
	\end{equation}
	
	The new coordinate system $\left(t,X,Y,\varphi\right)$ might be useful in some different calculations.
	
	\section{\label{sec:level2}Kretschmann scalar}
\begin{figure}[H]
		\centering
		\captionsetup{justification=justified}
		\begin{subfigure}{0.45\linewidth}
			\includegraphics[width=\linewidth]{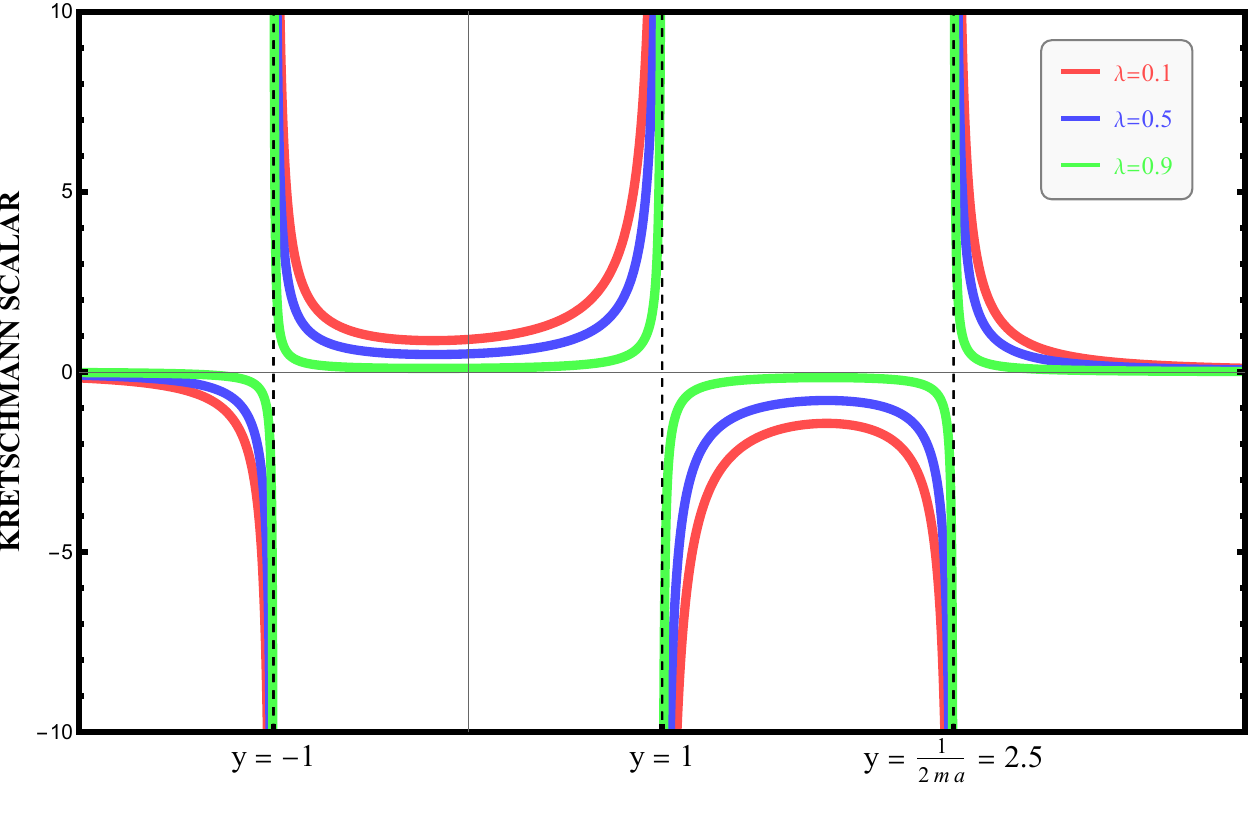}
			\caption{\it From Eq.~\eqref{xymetric}}
			\label{fig:subfig1}
		\end{subfigure}
		\begin{subfigure}{0.45\linewidth}
			\includegraphics[width=\linewidth]{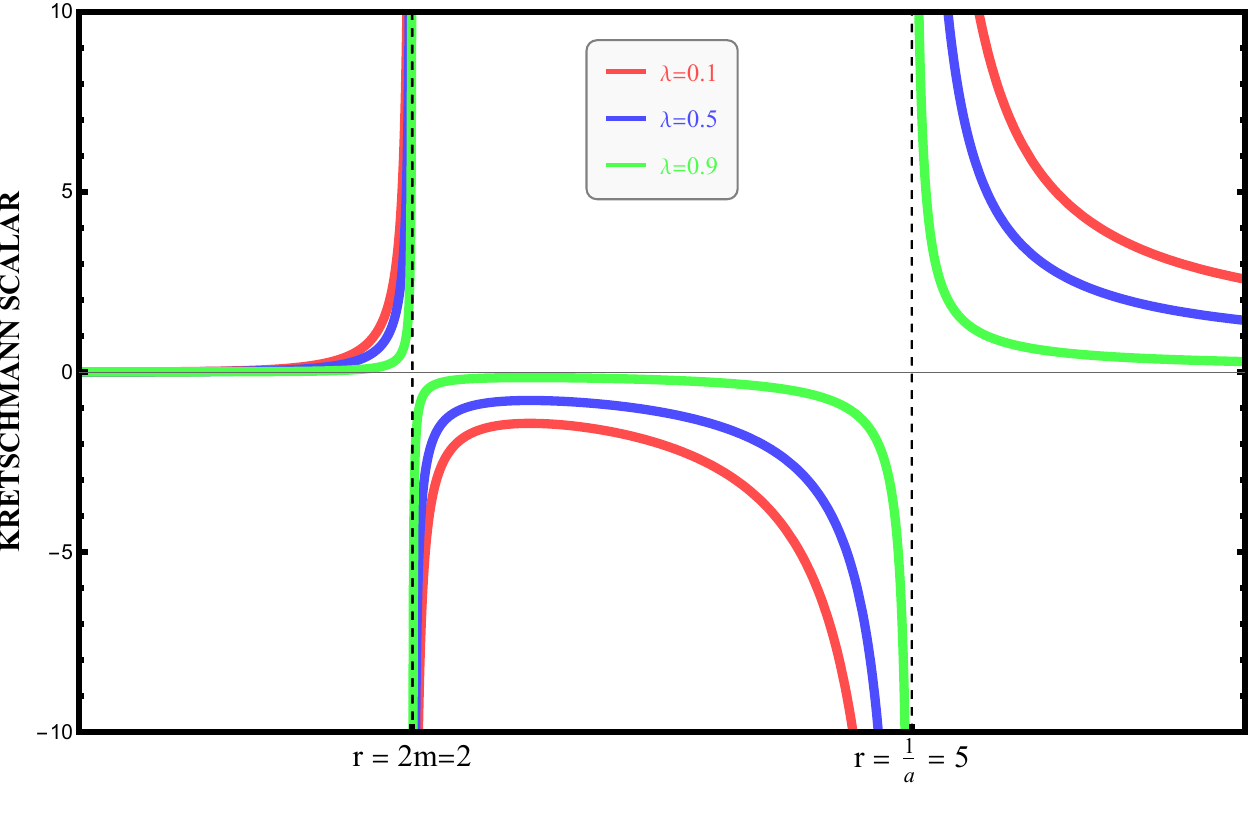}
			\caption{ From Eq.~\eqref{metric2}}
			\label{fig:subfig2}
		\end{subfigure}
		
		\captionsetup{justification=justified}
		\caption[]{\itshape (a): Kretschmann scalar as a function of $y$ at $x=0$, plotted for three values of $\lambda$ and for $m=1$ and $a=\dfrac{1}{5}$. The curvature singularities occur at $y=-1$, $y=1$ and $y=\frac{1}{2ma}$.\\
		(b): Kretschmann scalar as a function of $r$ at $\theta=\dfrac{\pi}{2}$, plotted for three values of $\lambda$ and for $m=1$ and $a=\dfrac{1}{5}$. The Kretschmann scalar exhibits singular behavior at $r=2m$ and $r=\frac{1}{a}=5$, indicating curvature singularities in the space-time.}
		\label{KRETSCHMANN SCALAR xy}
	\end{figure}

\begin{figure}[H]
		\centering
		\includegraphics[width=0.45\textwidth]{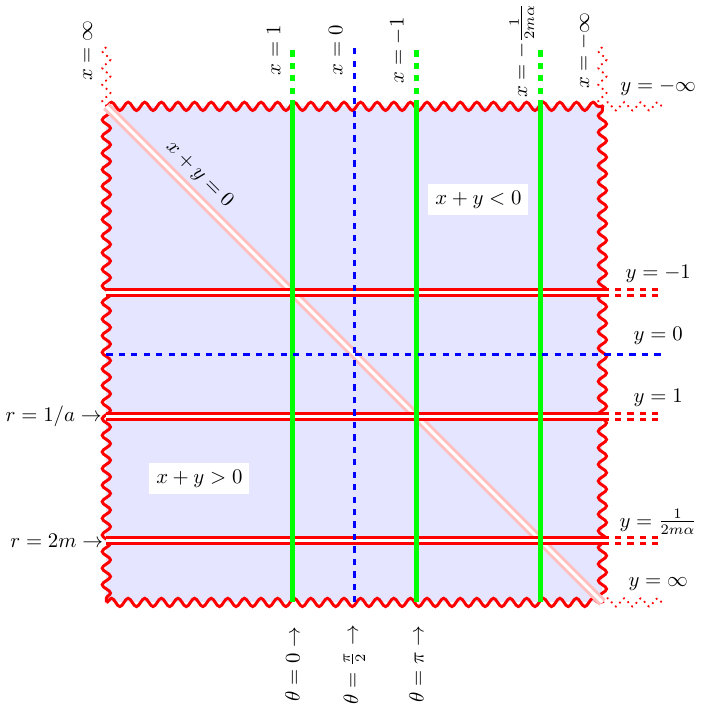}
		\caption{\itshape 
			Representation of the $xy$-coordinate plane based on the metric given in Eq.~\eqref{xymetric}. In this figure, the entire range of $x$ and $y$, i.e., $\left(-\infty,+\infty\right)$, is displayed. The coordinates $r$ and $\theta$, which are related to $x$ and $y$ via the transformations $y=-\dfrac{1}{a r}$ and $x=\cos\left(\theta\right)$, are also shown. The double red lines indicate singularities, the green lines mark the roots of the function $g\left(x\right)$, and the dashed blue lines correspond to $x=0$ and $y=0$.}
		\label{fig:C1}
	\end{figure}

\begin{figure}[H]
		\centering
		\captionsetup{justification=justified}
		\begin{subfigure}{0.493\linewidth}
			\includegraphics[width=\linewidth]{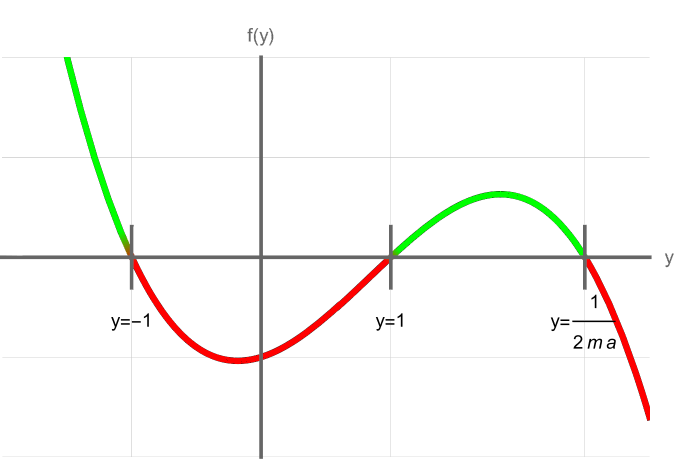}
			\caption{\it }
			\label{fig:subfig1}
		\end{subfigure}
		\begin{subfigure}{0.493\linewidth}
			\includegraphics[width=\linewidth]{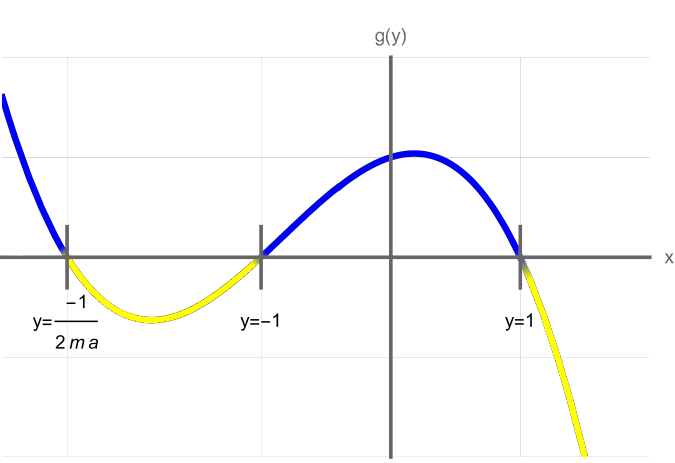}
			\caption{ }
			\label{fig:subfig2}
		\end{subfigure}
		
		\captionsetup{justification=justified}
		\caption[]{\itshape In part (a), the graph of the function $f\left(y\right)$ is plotted. The admissible regions, where $f\left(y\right)>0$, are shown in green, while the inadmissible regions, where $f\left(y\right)<0$, are shown in red. In part (b), the graph of the function $g\left(x\right)$ is displayed. The regions where $g\left(x\right)$ is positive are marked in blue, and the regions where $g\left(x\right)$ is negative are marked in yellow.}
		\label{fig:C33}
	\end{figure}

\begin{figure}[h]
		\centering
		\includegraphics[width=0.5\textwidth]{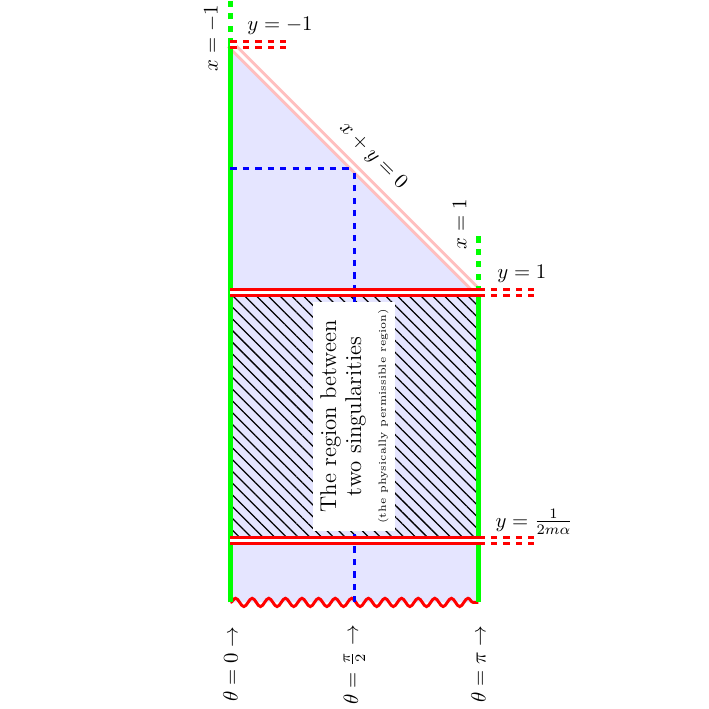}
		\caption[]{\it The shaded region represents the admissible domain in the $r$ and $\theta$ coordinate system from Eq.~\eqref{metric2}.}
		\label{fig:C4}
	\end{figure}

	One of the fundamental features of any metric is the behavior of its curvature invariants, as these invariants can indicate the presence of singularities. In this part, we compute and plot the Kretschmann scalar to analyze the singular structure of the accelerating FJNW spacetime.\\
	Since the original functions \(\mathcal{F}(Y)\) and \(\mathcal{G}(X)\) are cubic polynomials, finding their roots requires solving third-degree equations. However, after applying a suitable linear transformation and rescaling of $m$, $a$, $t$ and $\varphi$, we can fix two of the roots at $1$ and $-1$ and reach to the following simple form
	\begin{equation}
		\label{xyfg}
		\begin{split}
			&\mathcal{F}\left(Y\right)\rightarrow f\left(y\right)=-\left(1-y^{2}\right)\left(1-2a m y\right),\\
			&\mathcal{G}\left(X\right)\rightarrow g\left(x\right)=\left(1-x^{2}\right)\left(1+2a m x\right),
		\end{split}
	\end{equation}
	so
	\begin{equation}
		\label{xymetric}
		\begin{aligned}
			&ds^{2}=-\dfrac{f\left(y\right)^{\lambda}}{a^{2}\left(x+y\right)^{2\lambda}}d\tau^{2}+ \\ &\dfrac{1}{a^{2}\left(x+y\right)^{2\left(2-\lambda\right)}}\left[\dfrac{dy^{2}}{f\left(y\right)^{\lambda}}+f\left(y\right)^{1-\lambda}\left(\dfrac{dx^{2}}{g\left(x\right)}+\dfrac{d\phi^{2}}{g\left(x\right)^{-1}}\right)\right].
		\end{aligned}
	\end{equation}\\
	This form of metric is easier to work with because here roots of $f\left(y\right)$ and $g\left(x\right)$ are apparent from their form and there is no need to explicitly solve a cubic equation. To determine regions in this spacetime where there is genuine singular behavior we first look at regions of the spacetime where coefficients of the metric are zero or divergent. This gives us regions defined by $x+y=0$, $y = \pm1$, $y = \dfrac{1}{2a m}$, $x = \pm1$, and $x = -\dfrac{1}{2a m}$. However, these regions are not necessarily singular; therefore, to determine which of these regions are genuinely  singular, we calculate Kretschmann scalar, and see its behavior near the regions determined above. \\
	Specifically, we have plotted the Kretschmann scalar for the metric in Eq.~\eqref{xymetric} as a function of $y$ for some fixed values of $\lambda$ and $x$ in part (a) of FIG.~(\ref{KRETSCHMANN SCALAR xy}). For metric in Eq.~\eqref{metric2}, the Kretschmann scalar is depicted as a function of $r$ for some fixed values of $\lambda$ and $\theta$ in part (b) of FIG.~(\ref{KRETSCHMANN SCALAR xy}).\\
	As shown in the part (a) of FIG.~(\ref{KRETSCHMANN SCALAR xy}), the Kretschmann scalar diverges at three specific points $y=-1$, $y=1$ and $y=\frac{1}{2ma}$, and shown in the part (b) of FIG.~(\ref{KRETSCHMANN SCALAR xy})(b), the scalar diverges at two specific points $r=2m$ and $r=\frac{1}{a}$. These divergences indicate the presence of true singularities in these regions.\\
	Since the Ricci scalar has a simpler analytic form compared to the Kretschmann scalar, we proceed to compute it for metric in Eq.~\eqref{xymetric} as follows\\
	\begin{equation}
		\label{richiixy}
		\begin{split}
			R=&\dfrac{8a^{2}\left(\lambda^{2}-1\right)\left(x+y\right)^{4-2\lambda}}{\left(y^{2}-1\right)^{2-\lambda}\left(1-2may\right)}\left[\ \dfrac{ma\left(xy^{2}-x+2y\right)}{2}
			\right.\\
			&\left. \dfrac{1}{4}+m^{2}a^{2}\left(xy^{3}+\dfrac{1}{4}y^{4}-xy+\dfrac{1}{2}y^{2}+\dfrac{1}{4}\right)\right].\\ 
		\end{split}
	\end{equation}
The Ricci scalar corresponding to accelerating FJNW metric in Eq.~\eqref{metric2} is written as follows
\begin{equation}
		\label{richiir}
		\begin{split}
			R=&\dfrac{2\left(\lambda^{2}-1\right)\left(1-\dfrac{2m}{r}\right)^{\lambda}\left(1+ar\cos\left(\theta\right)\right)^{4-2\lambda}}{r^{2}\left(r-2m\right)^{2}}\\
			&\left[\ -4a^{2}r^{3}+a^{2}r^{4}+\left(m+ma^{2}r^{2}\right)^{2}
			\right.\\
			&\left. -2amr\left(r-2m\right)\left(a^{2}r^{2}-1\right)\cos\left(\theta\right)\right].\\ 
		\end{split}
	\end{equation}
	\\
	At this stage, we aim to determine the admissible regions for the variables $x$, $y$ and $r$. As a first step, we assume that these variables are defined over the entire $\left(-\infty,+\infty\right)$. Based on this assumption, the full two-dimensional $x$ and $y$ plane is plotted in  FIG. (\ref{fig:C1}) to establish the initial domain of the coordinate space under consideration.
	Turning to the FIG. (\ref{fig:C1}), we realize that for some specific $y$ there is singularity throughout the space, and we will see further that this will lead to very interesting and different results compared to the usual C-metric.\\
	As shown in FIG. (\ref{fig:C1}), the upper part of this line corresponds to $\left(x+y<0\right)$ and the lower part corresponds to $\left(x+y>0\right)$. Therefore, we clearly see in the relation $\dfrac{1}{\left(x+y\right)^{2\lambda}}$ that for some specific values of $\lambda$, this expression becomes imaginary and the upper part of the diagram no longer has a physical meaning, or at least it does not have a physical meaning in its conventional form.\\
		\begin{figure}[H]
		\centering
		\includegraphics[width=0.47\textwidth]{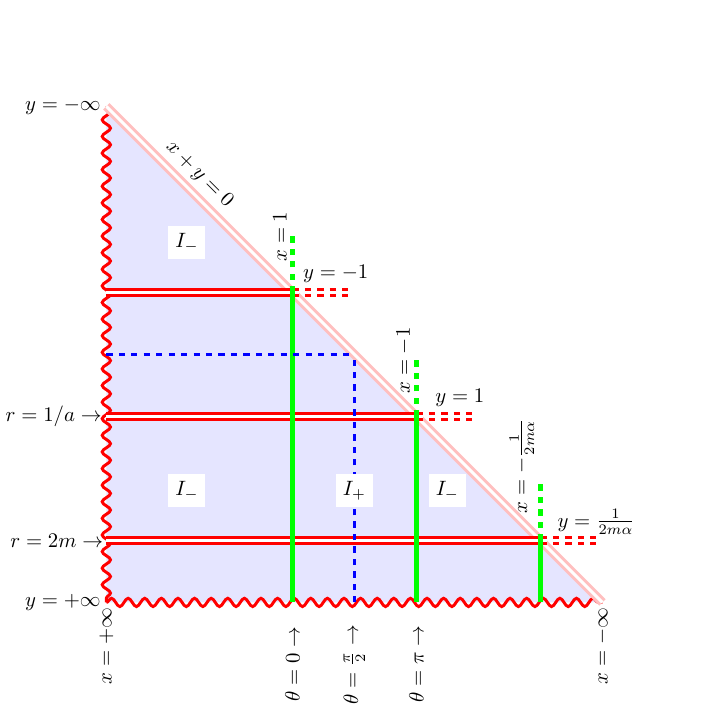}
		\caption[]{\it The set $I$ represents the admissible region. The subset  $I_{+}$corresponds to the part of $I$ where the function $g\left(x\right)$ is positive, and in this region, $\tau$ plays the role of time. Similarly, $I_{-}$	denotes the admissible region where $g\left(x\right)$ is negative, representing the domain in which the variable $y$ plays the role of time.}
		\label{fig:C2}
	\end{figure}
	To determine the admissible regions in FIG.~(\ref{fig:C2}), it is necessary to identify the areas where the function $f\left(y\right)$ is positive. In regions where $f\left(y\right)<0$, the function becomes inadmissible due to its appearance in the exponent $\lambda$. Therefore, to enable a more precise analysis, the graph of $f\left(y\right)$ is plotted in FIG.~(\ref{fig:C33}).\\
	The admissible region for the parameters is constrained by the requirement that $f\left(y\right)$ remains positive.\\
	We now simply display all admissible regions for Eq.~\eqref{metric2} in the $r$ and $\theta$ coordinates in  FIG. (\ref{fig:C4})
	\\
	
	\section{\label{sec:level2}effective potential and Geodesic Equations}
	In the absence of any force other than gravity, free particles move on geodesic paths. Therefore, calculating and checking these geodesics is one of the most important tasks for checking the structure of a space-time. Mathematically, to obtain these geodesies, we introduce the following action of the test particle in that space \cite{potential}.\\
	Here we assume that the motion of the test particles is governed by the following action:
	\begin{equation}
		\label{v1}
		\begin{split}
			&\mathcal{S}\equiv\int d\lambda\left(g_{\mu\nu}\dfrac{d\dot{x}^{\mu}}{d\lambda}\dfrac{d\dot{x}^{\nu}}{d\lambda}\right),
		\end{split}
	\end{equation}
	According to Eq.~\eqref{v1} the Lagrangian is written as follows:
	\begin{equation}
		\label{v2}
		\begin{split}
			&\mathcal{L}=\dfrac{1}{2}g_{\mu\nu}\dot{x}^{\mu}\dot{x}^{\nu}.
		\end{split}
	\end{equation}
	By variation of the action, the well-known Euler-Lagrange equation can be written in the following form:
	\begin{equation}
		\label{v3}
		\begin{split}
			&\dfrac{d}{d\lambda}\dfrac{\partial\mathcal{L}}{\partial\dot{x}^{\mu}}-\dfrac{\partial\mathcal{L}}{\partial{x}^{\mu}}=0.
		\end{split}
	\end{equation}
	The use of four-momentum allows us to simultaneously calculate geodesics for both massive and massless particles. Therefore, we define four-momentum as follows:
	\begin{equation}
		\label{v4}
		\begin{split}
			&p^{\mu}=\dfrac{dx^{\mu}}{d\lambda}\equiv\dot{x}^{\mu}
		\end{split}
	\end{equation}
A point to be noted here is that $\lambda$ is an affine parameter and more precisely it can be calculated as the proper time of a particle with rest mass $m_{0}$ along the geodesic.\\
	Therefore, using from the Lagrangian in Eq.~\eqref{v2}, the four-momentum can be written as follows
	\begin{equation}
		\label{v5}
		\begin{split}
			&p_{\mu}=\dfrac{\partial\mathcal{L}}{\partial\dot{x}^{\mu}}=\dot{x}_{\mu}=g_{\mu\nu}\dot{x}^{\nu}.
		\end{split}
	\end{equation}
	We will rewrite Lagrangian in terms of the metric components in Eq.~\eqref{metric2} as follows:
	\begin{equation}
		\label{v7}
		\begin{split}
			2\mathcal{L}&=g_{\mu\nu}\dot{x}^{\mu}\dot{x}^{\nu}=-\dfrac{f(r)^\lambda h(r)^\lambda }{\Omega^{2\lambda}} \times\dot{t}^{2}	\\
			&+\dfrac{1}{\Omega^{2\left(2-\lambda\right)}}\times\dfrac{1}{f(r)^\lambda h(r)^\lambda }\times\dot{r}^{2} \\
			&+\dfrac{f(r)^{1-\lambda} h(r)^{1-\lambda} }{\Omega^{2\left(2-\lambda\right)}}\times\dfrac{r^{2}}{\mathcal{P}\left(\theta\right)}\times\dot{\theta}^{2} \\
			&+\dfrac{f(r)^{1-\lambda} h(r)^{1-\lambda}}{\Omega^{2\left(2-\lambda\right)}}\times\left(\mathcal{P}\left(\theta\right)r^{2}\sin\left(\theta\right)^{2}\right)\times\dot{\varphi}^{2}.
		\end{split}
	\end{equation}
	The four-momentum components are easily written as follows
	\begin{equation}
		\label{v8}
		\begin{split}
			&p_{t}=\dfrac{\partial\mathcal{L}}{\partial\dot{t}}=-\dfrac{f(r)^\lambda h(r)^\lambda}{\Omega^{2\lambda}}\dot{t},
		\end{split}
	\end{equation}
	\begin{equation}
		\label{v9}
		\begin{split}
			&p_{r}=\dfrac{\partial\mathcal{L}}{\partial\dot{r}}=\dfrac{1}{\Omega^{2\left(2-\lambda\right)}}\times\dfrac{1}{f(r)^\lambda h(r)^\lambda}\dot{r},
		\end{split}
	\end{equation}
	\begin{equation}
		\label{v10}
		\begin{split}
			&p_{\theta}=\dfrac{\partial\mathcal{L}}{\partial\dot{\theta}}=\dfrac{f(r)^{1-\lambda} h(r)^{1-\lambda} }{\Omega^{2\left(2-\lambda\right)}}\times\dfrac{r^{2}}{\mathcal{G}\left(\theta\right)}\dot{\theta},
		\end{split}
	\end{equation}
	and
	\begin{equation}
		\label{v11}
		\begin{split}
			&p_{\varphi}=\dfrac{\partial\mathcal{L}}{\partial\dot{\varphi}}=\dfrac{f(r)^{1-\lambda} h(r)^{1-\lambda} }{\Omega^{2\left(2-\lambda\right)}}\times\left(\mathcal{G}\left(\theta\right)r^{2}\sin\left(\theta\right)^{2}\right)\dot{\varphi}.
		\end{split}
	\end{equation}
	The Hamiltonian can be defined by the Legendre transformation of the Lagrangian,
	\begin{equation}
		\label{v12}
		\begin{split}
			&\mathcal{H}=p_{\mu}\dot{x}^{\mu}-\mathcal{L}=g_{\mu\nu}\dot{x}^{\mu}\dot{x}^{\nu}-\mathcal{L}.
		\end{split}
	\end{equation}
	It is quite easy to check that for the Lagrangian in Eq.~\eqref{v2} one has $\mathcal{H}=\mathcal{L}$.
	Hamilton's equations corresponding to this Hamiltonian can now be written as follows:
	\begin{equation}
		\label{100}
		\begin{split}
			&\dfrac{dp_{\mu}}{d\lambda}=\dfrac{d}{d\lambda}\left(\dfrac{\partial\mathcal{L}}{\partial\dot{x}^{\mu}}\right).
		\end{split}
	\end{equation}
	Since the metric and hence the Hamiltonian are not functions of $t$ and $\varphi$, there exist two conserved quantities associated with the conjugate momenta of $t$ and $\varphi$.\\
	The explicit form of these two conserved momenta is given below:
	\begin{equation}
		\label{101}
		\begin{split}
			&-p_{t}=\dfrac{f(r)^\lambda h(r)^\lambda}{\Omega\left(r,\theta\right)^{2\lambda}}\times\dot{t}=E
		\end{split},
	\end{equation}
	\begin{equation}
		\label{102}
		\begin{split}
			&p_{\varphi}=\dfrac{f(r)^{1-\lambda}h(r)^{1-\lambda} }{\Omega\left(r,\theta\right)^{2\left(2-\lambda\right)}}\times\left(\mathcal{G}\left(\theta\right)r^{2}\sin\left(\theta\right)^{2}\right)\times\dot{\varphi}=J.
		\end{split}
	\end{equation}
	Here, $J$ and $E$ may be interpreted as the angular momentum and energy of the test particle, respectively.\\
	We proceed to present the explicit expressions for the remaining Hamilton’s equations in the $r$ and $\theta$ components.\\
	Due to the metric dependence on $r$ and $\theta$, there is no constant of motion in this case and their equations are written as follows:
	\begin{equation}
		\label{64}
		\begin{aligned}
			&\dfrac{dp_{r}}{d\lambda}=\dfrac{\partial\mathcal{L}}{\partial r}=\lambda g_{tt}\left(\dfrac{h(r)_{,r}}{h(r)}+\dfrac{f(r)_{,r}}{f(r)}-2\dfrac{\Omega_{,r}}{\Omega}\right)\dot{t}^{2} \\
			&-\lambda g_{rr}\left(\dfrac{h(r)_{,r}}{h(r)}+\dfrac{f(r)_{,r}}{f(r)}+2\left(\dfrac{2}{\lambda}-1\right)\dfrac{\Omega_{,r}}{\Omega}\right)\times\dot{r}^{2} \\
			&+\left(1-\lambda\right)g_{\theta\theta}\left(\dfrac{h(r)_{,r}}{h(r)}+\dfrac{f(r)_{,r}}{f(r)}+\dfrac{2\left(\lambda-2\right)}{1-\lambda}\dfrac{\Omega_{,r}}{\Omega}\right)\times\dot{\theta}^{2}\\
			&+\left(1-\lambda\right)g_{\varphi\varphi}\left(\dfrac{h(r)_{,r}}{h(r)}+\dfrac{f(r)_{,r}}{f(r)}+\dfrac{2\left(\lambda-2\right)}{1-\lambda}\dfrac{\Omega_{,r}}{\Omega}\right)\times\dot{\varphi}^{2}\\
			&+\dfrac{2}{r}g_{\theta\theta}\times\dot{\theta}^{2}+\dfrac{2}{r}g_{\varphi\varphi}\times\dot{\varphi}^{2},\\
		\end{aligned}
	\end{equation}\\
where comma means partial derivative with respect to the coordinate and 
	\begin{equation}
		\label{65}
		\begin{split}
			\dfrac{dp_{\theta}}{d\lambda}&=\dfrac{\partial\mathcal{L}}{\partial \theta}=-2\lambda g_{tt}\dfrac{\Omega_{,\theta}}{\Omega}\times\dot{t}^{2}-2\left(2-\lambda\right)g_{rr}\dfrac{\Omega_{,\theta}}{\Omega}\times\dot{r}^{2} \\
			&-g_{\theta\theta}\left(2\left(2-\lambda\right)\dfrac{\Omega_{,\theta}}{\Omega}+\dfrac{\mathcal{G}_{,\theta}}{\mathcal{G}}\right)\times\dot{\theta}^{2}\\
			&+g_{\varphi\varphi}\left(\dfrac{2\left(\lambda-2\right)\Omega_{,\theta}}{\Omega}+\dfrac{\mathcal{G}_{,\theta}}{\mathcal{G}}+2\cot\left(\theta\right)\right)\dot{\varphi}.\\
		\end{split}
	\end{equation}
	Solving this coupled system of nonlinear equations analytically is challenging. However, by exploiting the conserved quantities associated with the test particle and restricting the motion to $\theta=\dfrac{\pi}{2}$, we can achieve significant simplifications. We begin by noting that the Lagrangian in Eq.~\eqref{v2} does not depend explicitly on $\lambda$; consequently, the Hamiltonian and therefore the following quantity is conserved and independent of $\lambda$:\\
	\begin{equation}
		\label{200}
		\begin{split}
			&\sigma=-2\mathcal{L}=-\left(g_{tt}\dot{t}^{2}+g_{rr}\dot{r}^{2}+g_{\theta\theta}\dot{\theta}^{2}+g_{\varphi\varphi}\dot{\varphi}^{2}\right)|_{\theta=\frac{\pi}{2}}.
		\end{split}
	\end{equation}
	This expression for $\sigma=1$ corresponds to massive test particles, while $\sigma=0$ applies to massless test particles. Since the motion is restricted to the subspace defined by $\theta=\dfrac{\pi}{2}$, we have $\dot{\theta}=0$. Utilizing the relations presented in Eq.~\eqref{101} and Eq.~\eqref{102}, we can express the quantities $\dot{t}$ and $\dot{\phi}$ in Eq.~\eqref{200} in the following form:
	\begin{equation}
		\label{201}
		\begin{split}
			&\dot{t}=\dfrac{E}{g_{tt}}|_{_{\theta=\frac{\pi}{2}}}=\dfrac{E}{f\left(r\right)^{\lambda}h\left(r\right)^{\lambda}},
		\end{split}
	\end{equation}
	and
	\begin{equation}
		\label{202}
		\begin{split}
			&\dot{\varphi}=\dfrac{J}{g_{\varphi\varphi}}|_{_{\theta=\frac{\pi}{2}}}=\dfrac{J}{r^{2}f\left(r\right)^{1-\lambda}h\left(r\right)^{1-\lambda}}.
		\end{split}
	\end{equation}
	By substituting Eq.~\eqref{201} and Eq.~\eqref{202} into Eq.~\eqref{200}, we arrive at the following equation:
	\begin{equation}
		\label{204}
		\begin{split}
			&\dfrac{E^{2}}{f\left(r\right)^{\lambda}h\left(r\right)^{\lambda}}-\dfrac{\dot{r}^2}{f\left(r\right)^{\lambda}h\left(r\right)^{\lambda}}-\dfrac{J^{2}}{r^{2}f\left(r\right)^{1-\lambda}h\left(r\right)^{1-\lambda}}=\sigma
		\end{split}
	\end{equation}
	Using \eqref{204}, we can write the effective potential and geodesic equation of the test particles.
	\subsection{Effective Potential}
\begin{figure}[H]
		\centering
		\includegraphics[width=0.47\textwidth]{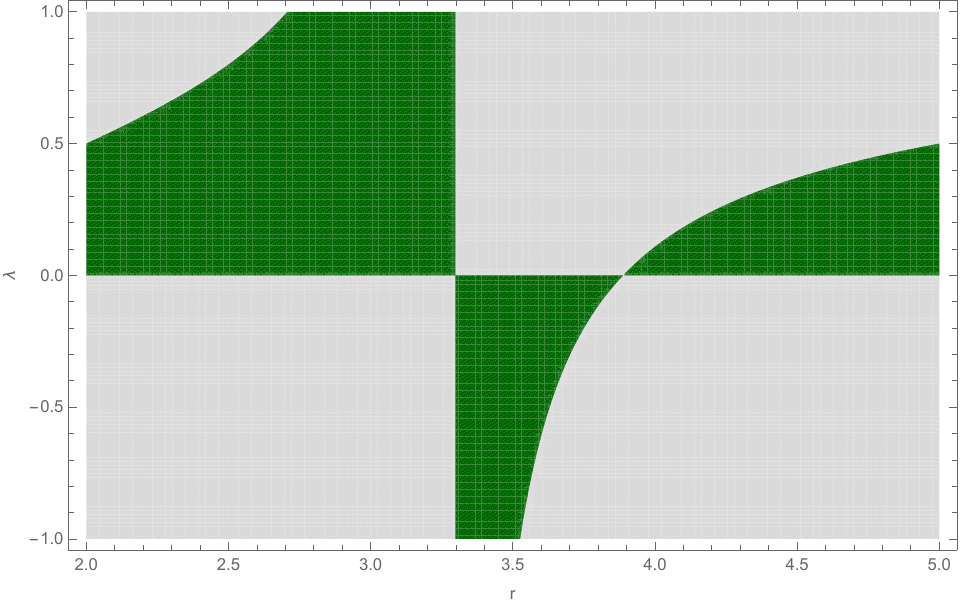}
		\caption[]{\it  Circular orbits in the plane $\left(r,\lambda\right)$ for $m=1$ and $a=\dfrac{1}{5}$. Green regions correspond to values of $r$ and $\lambda$ where circular orbits are allowed, while grey areas indicate where such orbits do not exist.}
		\label{J1}
	\end{figure}

	By examining Eq.~\eqref{204}, we observe that it can be algebraically rearranged to yield an equivalent form, given by 
	\begin{equation}
		\label{205}
		\begin{split}
			&V_{eff}=\dot{r}^{2},
		\end{split}
	\end{equation}
	where
	\begin{equation}
		\label{206}
		\begin{aligned}
			V_{eff}=&f\left(r\right)^{\lambda}h\left(r\right)^{\lambda}\\
			&\times\left[-\sigma+\dfrac{E^{2}}{f\left(r\right)^{\lambda}h\left(r\right)^{\lambda}}-\dfrac{J^{2}}{r^{2}f\left(r\right)^{1-\lambda}h\left(r\right)^{1-\lambda}}\right]
		\end{aligned}\\
	\end{equation}
	This structure closely resembles the energy equation of a classical one-dimensional particle in Newtonian mechanics, though it is not identical. Eq.~\eqref{206} can be written in the following form
	\begin{equation}
		\label{208}
		\begin{split}
			&\dfrac{1}{2}\dot{r}^{2}=\mathcal{E}-\dfrac{f\left(r\right)^{\lambda}h\left(r\right)^{\lambda}}{2}\left[\sigma+\dfrac{J^{2}}{r^{2}f\left(r\right)^{1-\lambda}h\left(r\right)^{1-\lambda}}\right],
		\end{split}
	\end{equation} 
where $\mathcal{E}=\dfrac{E^{2}}{2}$. Hence, the test particle effectively behaves like a classical particle moving in an effective potential given by
	\begin{equation}
		\label{209}
		\begin{split}
			&\mathcal{V}_{eff}\left(r\right)=\dfrac{f\left(r\right)^{\lambda}h\left(r\right)^{\lambda}}{2}\left[\sigma+\dfrac{J^{2}}{r^{2}f\left(r\right)^{1-\lambda}h\left(r\right)^{1-\lambda}}\right].
		\end{split}
	\end{equation} 
	This analogy allows us to employ standard techniques from Newtonian mechanics to analyze the motion of the test particle. For instance, circular orbits can be determined by solving equations
	\begin{equation}
		\label{210}
		\begin{split}
			\mathcal{V}_{eff}&=\mathcal{E},\\
			\frac{d\mathcal{V}_{eff}}{dr}&=0.
		\end{split}
	\end{equation}
	To identify the values of $J^{2}$ and $E$ that allow for circular orbits at a given radius $r$, we solve the governing equations by expressing $E$ and $J^{2}$ as functions of $r$. The resulting expressions are presented below:
	\begin{equation}
		\label{211}
		\begin{aligned}
			J^{2}=&\dfrac{-r^{2}\lambda f\left(r\right)^{1-\lambda}h\left(r\right)^{1-\lambda}\left(m+ma^{2}r^{2}-a^{2}r6{3}\right)}{m+2\lambda-r+ma^{2}r^{2}\left(2\lambda-3\right)-2a^{2}r^{3}\left(\lambda-1\right)},
		\end{aligned}
	\end{equation}
	and
	\begin{equation}
		\label{212}
		\begin{aligned}
			\mathcal{E}=&f\left(r\right)^{\lambda}h\left(r\right)^{\lambda}\\
			&\times\frac{m\left(1+\lambda+a^{2}r^{2}\left(\lambda-3\right)\right)-a^{2}r^{3}\left(\lambda-2\right)-r}{2m\left(1+2\lambda+a^{2}r^{2}\left(2\lambda-3\right)\right)-4a^{2}r^{3}\left(\lambda-1\right)-2r}.
		\end{aligned}
	\end{equation}\\
	Since $J^{2}$ must remain positive, we conclude from these solutions that circular orbits are only possible at those values of $r$ for which the expression for $J^{2}$ is positive. This leads to a condition that allows us to determine whether a circular orbit exists at a given radius $r$. FIG.~(\ref{J1}) illustrates the  corresponding values of parameters $r$ and $\lambda$ that permit circular orbits.\\
	Not all circular orbits are stable. To determine their stability, we evaluate $\dfrac{d^{2}\mathcal{V}_{eff}}{d^{2}r}$ at the radius of the circular orbit. If $\dfrac{d^{2}\mathcal{V}_{eff}}{d^{2}r}>0$, 
	the orbit is stable, meaning that small perturbations around it remain bounded. This gives us the stability condition for circular orbits.
	In total, two independent conditions must be satisfied for a stable circular orbit to exist:
	\begin{equation}
		\label{76}
		\begin{split}
			J^{2}&\geq0,\\
			\dfrac{d^{2}\mathcal{V}_{eff}}{d^{2}r}&>0.
		\end{split}
	\end{equation}
	
\begin{figure}[H]
		\centering
		\includegraphics[width=0.47\textwidth]{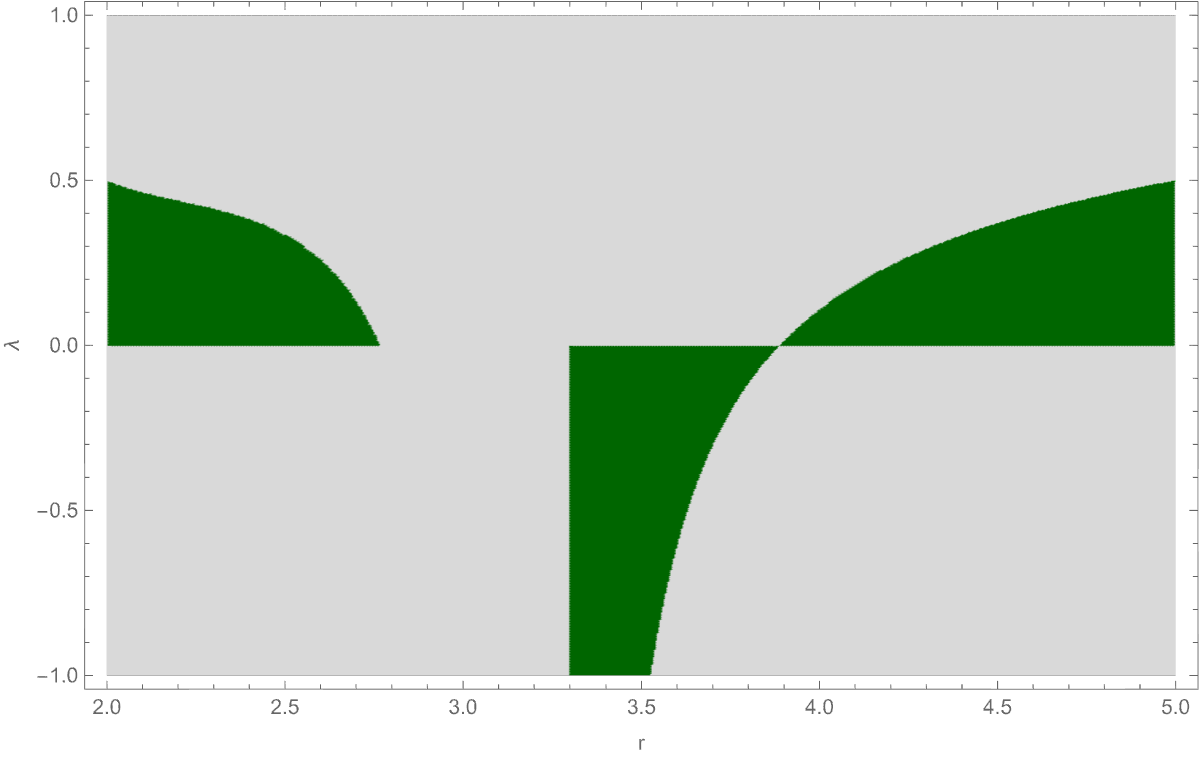}
		\caption[]{\it Green regions indicate points where both conditions $J^{2}\geq0$ and $\dfrac{d^{2}\mathcal{V}_{eff}}{d^{2}r}>0$ are satisfied, corresponding to the existence of stable circular orbits. In contrast, the grey areas mark regions where circular orbits either do not exist or are unstable.}
		\label{JJ}
	\end{figure}

\begin{figure}[H]
		\centering
		\includegraphics[width=0.47\textwidth]{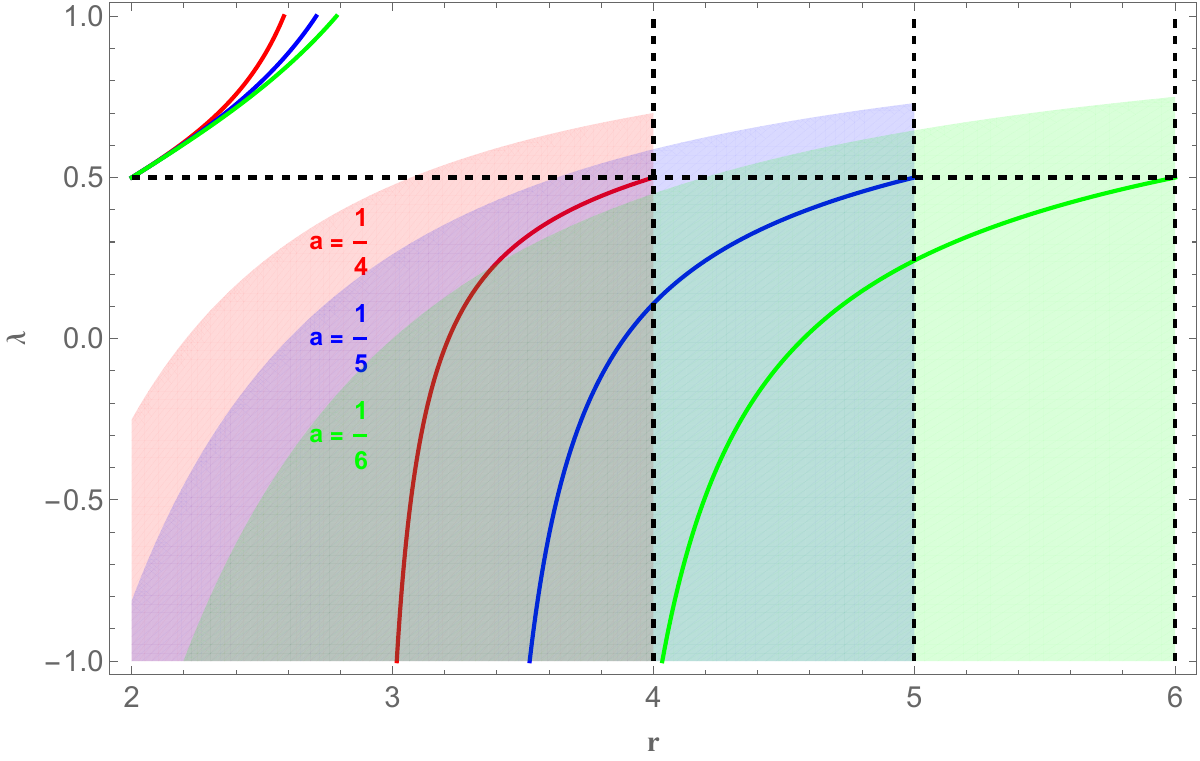}
		\caption[]{\it Colored lines indicate the locations where circular orbits for massless particles exist. The surrounding colored regions show where such orbits are stable. The plots are generated for different values of parameter $a$ and $m=1$. The overall structure of the regions admitting stable circular orbits appears to remain qualitatively unchanged as $a$ varies.}
		\label{J5}
	\end{figure}

	It can be shown that both conditions reduce to the requirement that two polynomials, say $p(r,\lambda)$ and $q(r,\lambda)$, are simultaneously positive: 
	\begin{equation}
		\label{76}
		\begin{split}
			p(r,\lambda)&>0,\\
			q(r,\lambda)&>0.
		\end{split}
	\end{equation}
	The explicit expressions for these polynomials are lengthy and not particularly insightful, so we do not present them here. However, we can still determine the values of $r$ and $\lambda$ for which stable circular orbits exist. These regions are shown in FIG.~(\ref{JJ})
	Up to this point, we have focused on massive particles. For massless particles, however, $\sigma=0$; as a result, the potential in Eq.~\eqref{209} simplifies accordingly to 
	\begin{equation}
		\label{219}
		\begin{split}
			&\mathcal{V}_{eff}\left(r\right)=\dfrac{J^{2}}{2r^{2}f\left(r\right)^{1-2\lambda}h\left(r\right)^{1-2\lambda}}.
		\end{split}
	\end{equation}
	To determine the values of $E$ and $J$ that allow for circular orbits at radius $r$, as in the massive case, we need to solve the following equations
	\begin{equation}
		\label{220}
		\begin{split}
			E=&\dfrac{J^{2}}{2r^{2}f\left(r\right)^{1-2\lambda}h\left(r\right)^{1-2\lambda}},\\
			\frac{d\mathcal{V}_{eff}}{dr}=&\dfrac{J^{2}}{r^{3}f\left(r\right)^{1-2\lambda}h\left(r\right)^{1-2\lambda}}\\
			&\times\left(rh\left(r\right)f\left(r\right)_{,r}\left(\lambda-\dfrac{1}{2}\right)\right.\\
			&\left.
			+rf\left(r\right)\left(h\left(r\right)_{,r}\left(1-\dfrac{1}{2}\right)-h\left(r\right)\right)\right).
		\end{split}
	\end{equation}
	Upon examining Eq.~\eqref{220}, we find that solutions only exist for specific, discrete values of $r$. Moreover, Eq.~\eqref{220} yields only the ratio of $\dfrac{E}{J}$, rather than individual values. These discrete radii correspond to the roots of the following polynomial
	\begin{equation}
		\label{221}
		\begin{split}
			H_{1}\left(r,\lambda\right)=&m\left(\lambda+\dfrac{1}{2}\right)-\dfrac{r}{2}+a^{2}m\left(\lambda-\dfrac{3}{2}\right)\\
			&-a^{2}r^{3}\left(\lambda-1\right).
		\end{split}
	\end{equation}
	As discussed earlier, the stability of these orbits can be assessed by evaluating $\frac{d\mathcal{V}_{eff}}{dr}>0$ condition at the orbit radius. For massless particles, this condition reduces to the simpler form
	
	\begin{equation}
		\label{222}
		\begin{split}
			H_{2}\left(r,\lambda\right)=&2ma^{2}r\left(\lambda-\dfrac{3}{2}\right)-3a^{2}r^{2}\left(\lambda-1\right)-\dfrac{1}{2}.
		\end{split}
	\end{equation}
	In summary, stable circular orbits exist for values of $\left(r,\lambda\right)$ that satisfy the following conditions
	\begin{equation}
		\label{210}
		\begin{split}
			H_{1}\left(r,\lambda\right)&=0,\\
			H_{2}\left(r,\lambda\right)&>0.
		\end{split}
	\end{equation}
	Fig.~\ref{J5} shows the regions of admissible values of $r$ and $\lambda$ where such stable circular orbits occur. 
	\subsection{Geodesic}
\begin{figure}[H]
		\centering
		\includegraphics[width=0.23\textwidth]{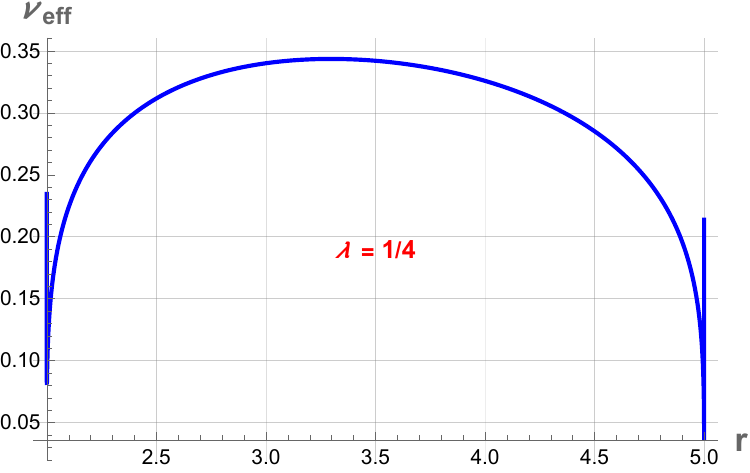}
		\includegraphics[width=0.2\textwidth]{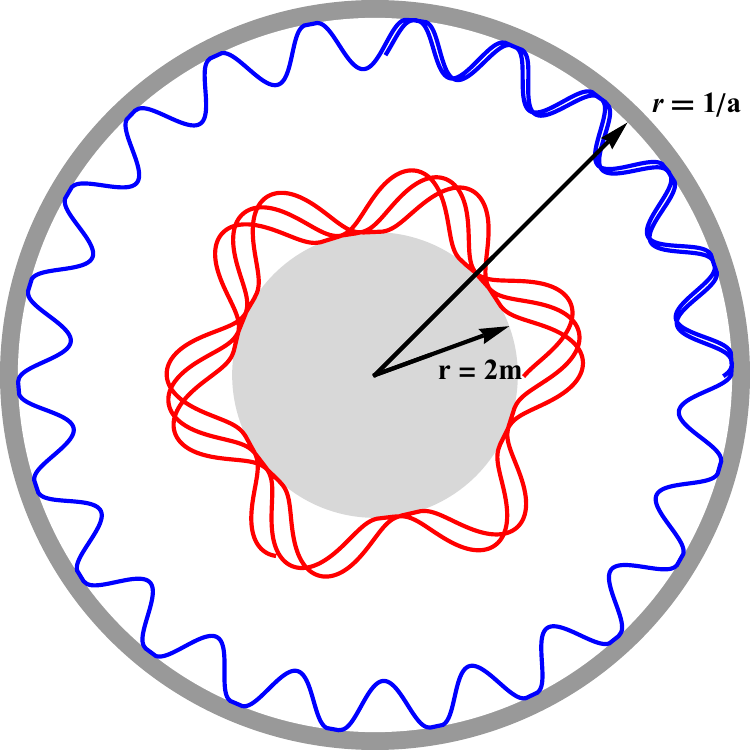}
		\includegraphics[width=0.23\textwidth]{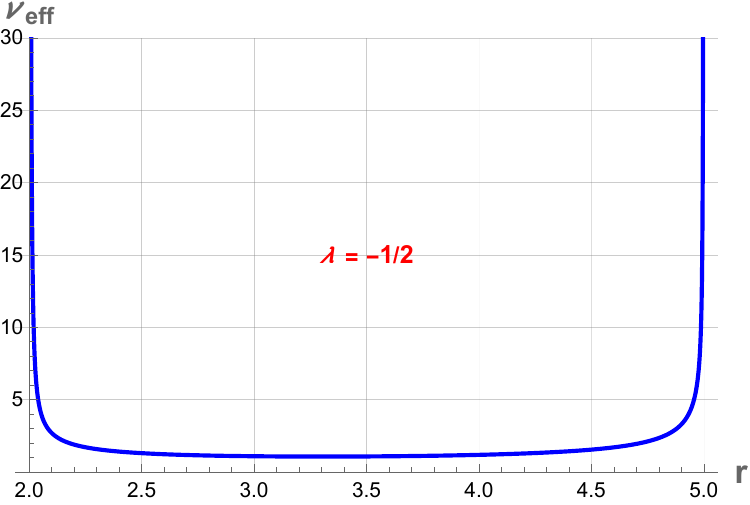}
		\includegraphics[width=0.2\textwidth]{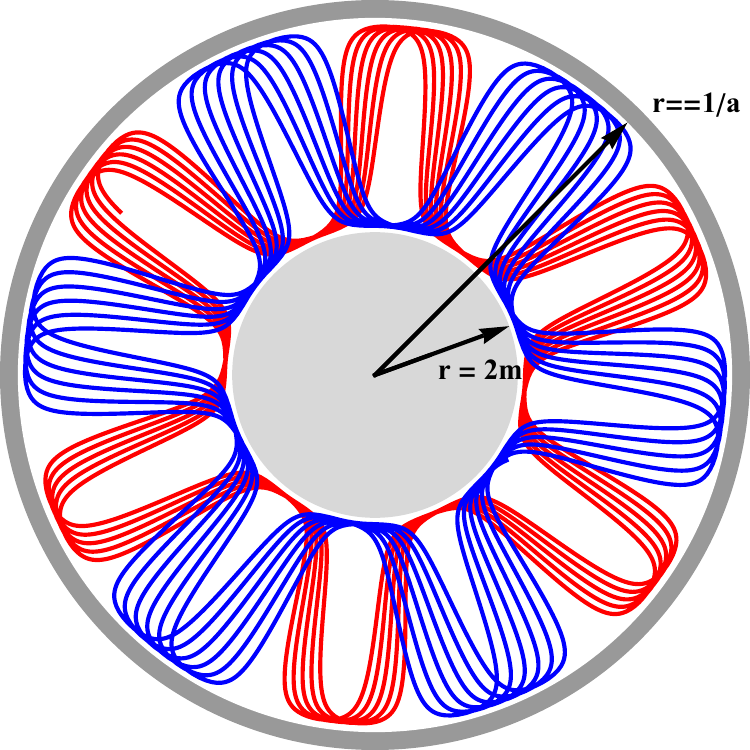}
		\includegraphics[width=0.23\textwidth]{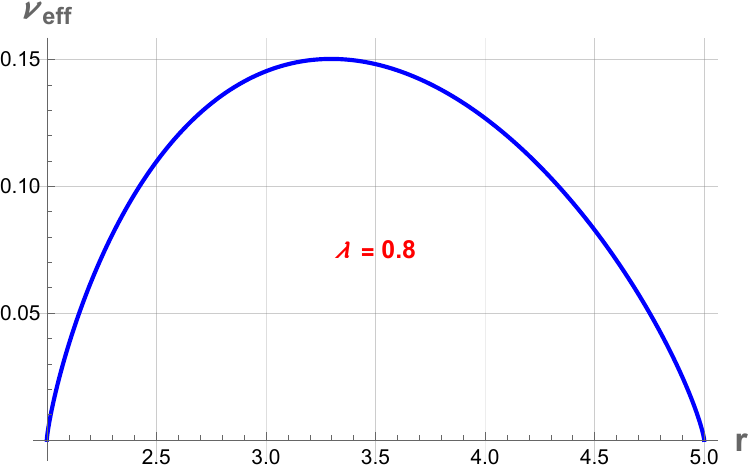}
		\includegraphics[width=0.2\textwidth]{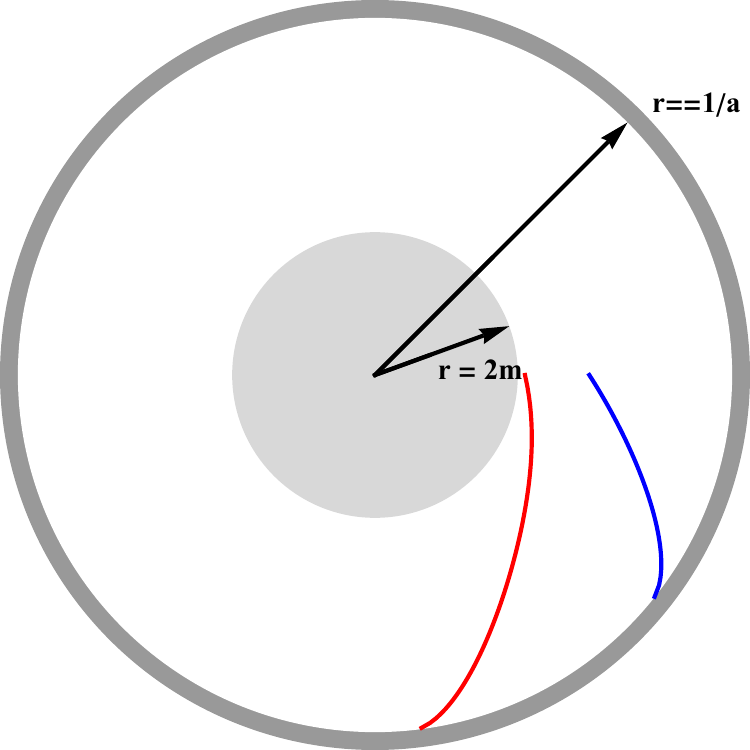}
		\caption[]{\it The effective potential is shown for three values of $\lambda=\dfrac{1}{4}$, $\lambda=-\dfrac{1}{2}$ and $\lambda=0.8$, each illustrating a distinct typical behavior. For each case, the corresponding geodesic trajectory obtained by solving the geodesic equations with the plotted potential is also shown. All other configurations exhibit qualitatively similar behavior to one of these three cases.}
		\label{GVp}
	\end{figure}

\begin{figure}[H]
		\centering
		\includegraphics[width=0.23\textwidth]{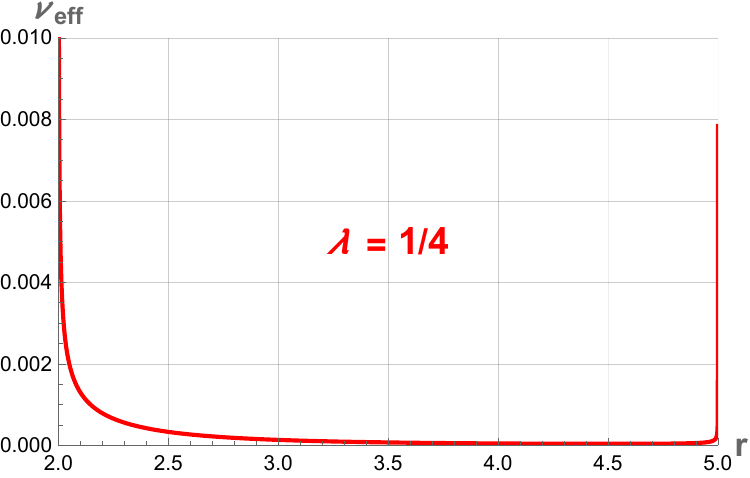}
		\includegraphics[width=0.2\textwidth]{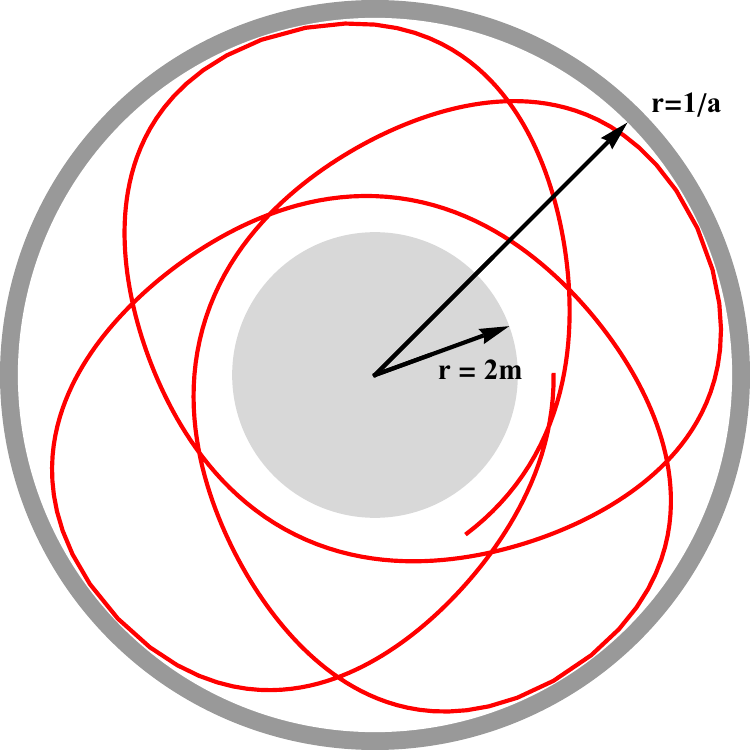}
		\includegraphics[width=0.23\textwidth]{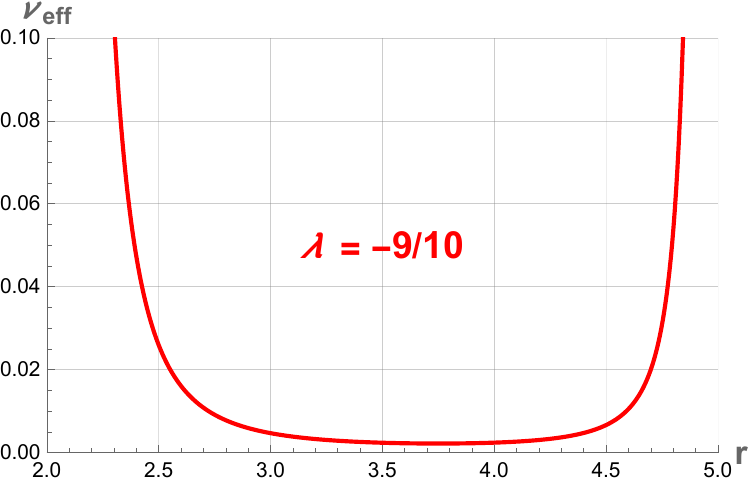}
		\includegraphics[width=0.2\textwidth]{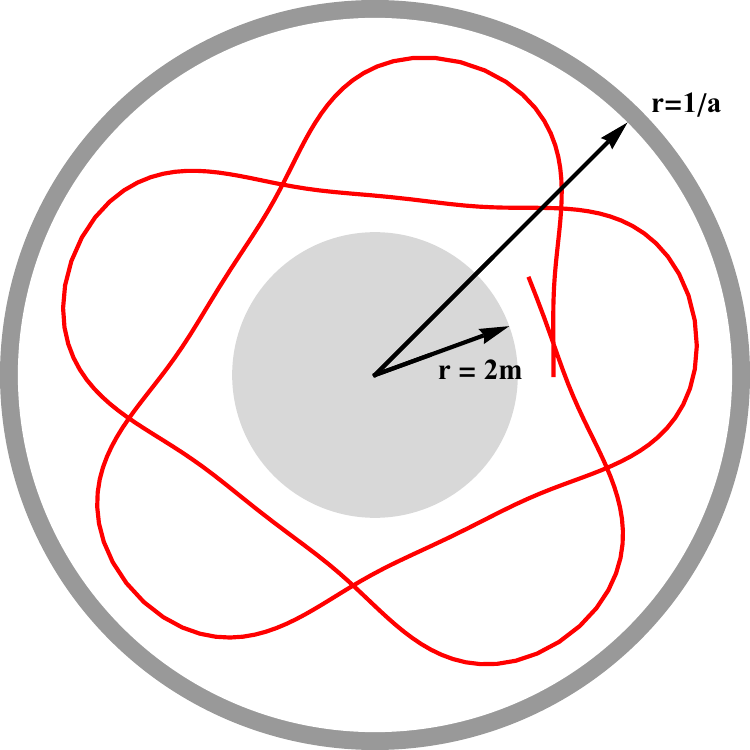}
		\includegraphics[width=0.23\textwidth]{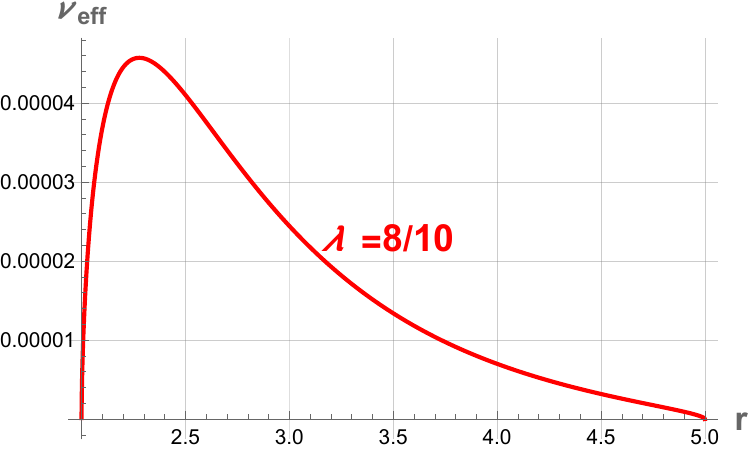}
		\includegraphics[width=0.2\textwidth]{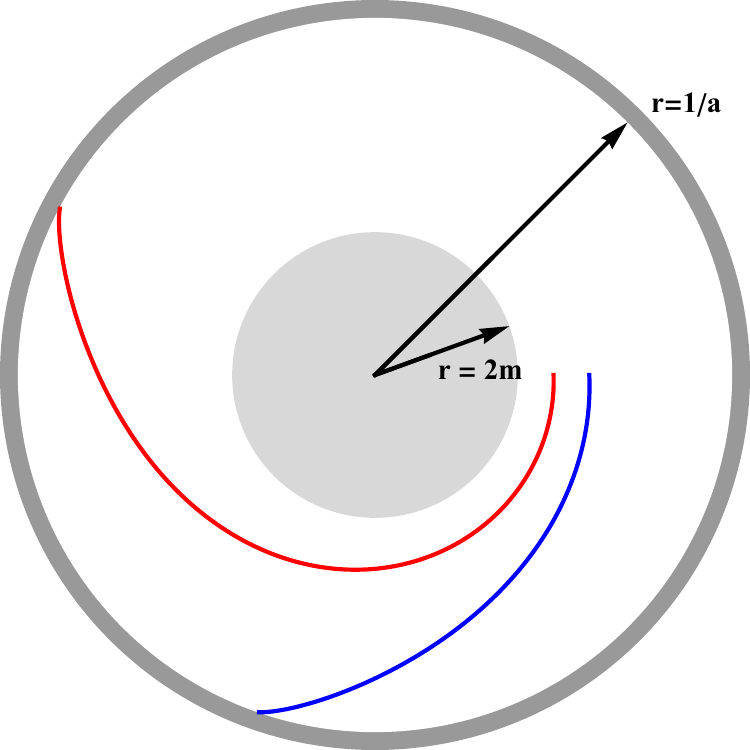}
		\caption[]{\it Effective potential for massless particles ($\sigma=0$) for $\lambda=\dfrac{1}{4}$, $\lambda=-\dfrac{9}{10}$, and $\lambda=\dfrac{8}{10}$.}
		\label{GVP}
	\end{figure}

	Up to this point, we have analyzed the behavior of the effective potential for test particles. We now turn our attention to the geodesics in the plane defined by equation $\theta=\dfrac{\pi}{2}$. Starting from Eq.~\eqref{208}, if we take its derivative and divide both sides by $\dfrac{dr}{d\lambda}$, we arrive at 
	\begin{equation}
		\label{502}
		\begin{split}
			&\dot{\varphi}=\dfrac{J\Omega\left(r,\theta\right)^{2\left(2-\lambda\right)} \mathcal{G}\left(\theta\right)r^{2}\sin\left(\theta\right)^{2}}{f(r)^{1-\lambda}h(r)^{1-\lambda}}.
		\end{split}
	\end{equation}
	To fully determine the geodesics in this plane, Eq.~\eqref{502} is also required. Thus, we are led to a system consisting of two second-order differential equations, given as
	\begin{equation}
		\label{225}
		\begin{aligned}
			\dfrac{dr^{2}}{d\lambda^{2}}=&\dfrac{\lambda mf\left(r\right)^{\lambda}h\left(r\right)^{\lambda}\left(1+r^{-2}f\left(r\right)^{\lambda-1}h\left(r\right)^{\lambda-1}\right)}{r\left(2m-r\right)}\\
			&-\lambda a^{2}r\left(f\left(r\right)^{\lambda}h\left(r\right)^{\lambda-1}-\dfrac{Jf\left(r\right)^{2\lambda-1}h\left(r\right)^{2\lambda-2}}{r^{2}}\right)\\
			&-\dfrac{Jf\left(r\right)^{2\lambda-2}h\left(r\right)^{2\lambda-2}}{denr^{3}}\\
			&\times\left(m\left(\lambda+1\right)-r+ma^{2}r^{2}\left(\lambda-3\right)-a^{2}r^{3}\left(\lambda-2\right)\right),
		\end{aligned}
	\end{equation} \\
	which must be solved to determine the paths of test particles.
	In general, solving this system analytically is not feasible. Therefore, we rely on numerical methods to explore and visualize the behavior of the system. In Fig.~\ref{GVp}, we present plots of the effective potential for several specific values of the parameters. These plots reveal three qualitatively distinct behaviors.\\
	First, for certain values of parameter $\lambda$, the effective potential lacks a minimum. In such cases, test particles inevitably fall into one of the two singularities, located at $r=2m$ or $r=\dfrac{1}{a}$, depending on the initial conditions.
	Second, when the potential exhibits a single minimum, it is observed that this minimum is relatively flat, with a small second derivative indicating a weak restoring force around the stable point.
	Third, for some values of parameter $\lambda$, the effective potential displays two distinct minima. In this case, the potential near each minimum is much steeper than in the previous scenario, reflected by a significantly larger second derivative and sharper curvature at the minima.
	Moreover, it is found that the activation energy required to escape the minimum near $\dfrac{1}{a}$ is substantially greater than that associated with the minimum near $r=2m$. These distinct types of behavior are consistent with the structure of stable circular orbits represented in Fig.~\ref{JJ}. In Fig.~\ref{GVp}, we also illustrate the trajectories of test particles as governed by the geodesic equations.\\
	We can carry out a similar analysis for massless particles by substituting $\sigma=0$ into the effective potential in Eq.~\eqref{209}. In this case, only two qualitatively distinct behaviors emerge, both in the form of the effective potential and in the resulting motion of the massless particles.
	The first behavior occurs for $\lambda>0.5$, where the effective potential lacks a minimum. As a result, depending on the initial conditions, the massless particle ultimately reaches one of the singularities located at $r=2m$ and $r=\dfrac{1}{a}$.
	The second behavior appears for $\lambda<\dfrac{1}{2}$. Here, the effective potential develops a single minimum, which is relatively flat indicating a weak restoring force near the stable point. Unlike the case of massive particles, no configuration is found in which the potential has two minima.
	These two characteristic forms of the effective potential, along with the corresponding particle trajectories, are illustrated in Fig.~\ref{GVP}.
	\section{CONCLUCION}
	In this work, we introduced a new class of exact solutions to the Einstein scalar field equations. We investigated curvature singularities of the class of accelerating spacetimes.	We also analyzed the motion of test particles in the class of accelerating metrics by computing the effective potential. This allowed us to identify the ranges of parameters $r$ and $\lambda$ that permit the existence of stable circular orbits. Our investigation revealed that, for massive particles, the effective potential exhibits three qualitatively distinct behaviors, while in the massless case, only two such behaviors are observed. It is interesting to explore quasi-normal modes of the class of metrics in the future.


	
	\nocite{apsrev41Control}
	\bibliographystyle{apsrev4-1}
	\bibliography{refcontrol,references.bib}

\end{document}